\documentclass{agujournal2019}

\pdfoutput=1
\draftfalse

\usepackage{amsmath}
\usepackage{bm}
\usepackage{url}

\begin{document}

\title{A new model for acoustic-poroelastic coupling of compressional body and Stoneley waves at a fault zone}

\authors{Shohei Minato\affil{1}, Tsutomu Kiguchi\affil{2}, and Ranajit Ghose\affil{1}}

 \affiliation{1}{Delft University of Technology}
 \affiliation{2}{Geological Survey of Japan, National Institute of Advanced Industrial Science and Technology (AIST)}

\affiliation{1}{Stevinweg 1, Delft, The Netherlands}
\affiliation{2}{Tsukuba Central 7, 1-1-1, Higashi, Tsukuba, Ibaraki, Japan}

\correspondingauthor{Shohei Minato}{s.minato-1@tudelft.nl}

\begin{keypoints}
\item We develop a new model that can estimate fluid pressure in an irregular borehole embedded in poroelastic media for an incident plane P wave
\item We derive analytical solutions for the new model and verify them by finite-difference numerical solutions of Biot poroelasticity equations
\item The modeled response at a fault zone is consistent with the observed data and reveals the heterogeneous permeability distribution
\end{keypoints}

\begin{abstract}
In vertical seismic profiling (VSP), Stoneley (tube) waves are generated due to the coupling between the borehole fluid and the surrounding poroelastic formation. The tube waves have been exploited in the past to infer the in-situ hydraulic properties. In order to understand better the physical mechanisms underlying the generation of tube waves at a fault zone, we develop a new model that calculates the pressure responses in a borehole. The model incorporates simultaneous effects of elastic impedance boundaries, fluid infiltration from poroelastic formation, and irregularities in the borehole radius. The analytical tube-wave amplitudes are derived from the new model assuming a normally incident plane P wave, verified by complete numerical solutions for Biot's theory of dynamic poroelasticity. We find that the upgoing and downgoing tube waves due to an elastic impedance boundary have opposite polarities, and those excited by a thin poroelastic layer have different wave shapes. The model also enables the prediction of a VSP response at a major fault zone in Japan (Nojima fault). Our quantitative evaluation suggests that tube waves are generated by elastic impedance boundaries and borehole irregularities around the main shear zone of the fault, as well as due to the presence of several porous layers. We also find that the modeled amplitudes agree well with the observation, especially when assuming a heterogeneous permeability distribution. The developed model and the presented results will be crucial in quantitatively interpreting the VSP data in order to estimate the fault zone's hydraulic properties.
\end{abstract}

\section*{Plain Language Summary} 
Estimating the hydraulic permeability of a fault zone has many applications, such as a better understanding of earthquake nucleation mechanisms and exploration of geothermal fields. For this purpose, a borehole is often drilled to the depth of a target fault zone. While a seismic wave travels, it deforms rocks and compresses water within them. At the same time, the water coming out of the rocks produces acoustic waves in the borehole. Analyzing such acoustic waves makes it possible to determine the in-situ permeability with low data acquisition costs. We have developed a new model to accurately calculate how the acoustic waves are produced and how they travel in a borehole located in a complex structure like a fault zone. Our model reveals that the acoustic waves have distinctive features when caused by different mechanisms, such as due to stiffness in the rocks, a fluid-bearing area, and an irregular borehole shape. Our calculation agrees well with data observed in the field at a fault zone in Japan. It also suggests that the permeability changes with depth. The new model helps us gain more insights into the acoustic wave data. It will enable accurate and efficient monitoring of the subsurface's permeability.

\section{Introduction}
Hydraulic properties of a fault zone, especially their spatially heterogeneous distribution, stress dependence, and temporal changes, are key to comprehending the fluid circulation in fractured media and the deformation processes in the upper crust. Measurements using boreholes drilled to the depth of a target fault zone have so far played a vital role in providing the hydraulic properties. For example, the scientific drilling of an active fault zone has revealed in the past the in-situ fracture geometry, fracture-induced seismic anisotropy, and depth varying stress orientations \cite<e.g.,>[among many others]{Ando2001,Zoback2010}. Monitoring the groundwater table in a borehole has enabled estimating the hydraulic permeability and its temporal changes at a fault zone that are correlated to remote earthquakes \cite{Xue2013}. Fluid injection experiments at a geothermal field have been exploited in order to constrain mechanisms associated with the hydro-mechanical response of fractured media \cite<e.g.,>[]{Amann2018}. In this vein, it is well known that the hydraulic permeability at a fault zone shows complex spatial variations depending on the architectures of the zone \cite{Faulkner2010}. Therefore, high-resolution information of the permeability distribution is vital to understand better the earthquake mechanisms and explore a geothermal field.

In contrast to controlled laboratory experiments, borehole measurements contain information of the hydraulic properties representing the in-situ condition (of stress and heterogeneity), thus without the effect of disturbance in the collected rock samples. One of the direct hydraulic experiments which generate in-situ fluid flow is the observation of acoustic waves in a fluid-filled borehole \cite<e.g.,>[]{Beydoun1985,Tang1996}. These measurements can efficiently provide hydraulic information at multiple depths compared with the packer tests that are focused on the properties within a specific depth interval \cite<e.g.,>[]{Cook2003}. The dynamic wavefield experiments observe the pressure perturbation in the borehole fluid surrounded by a porous formation. The wavefield includes the effects of the Biot slow wave due to dynamic poroelasticity \cite{Biot1956a,Biot1956b,Biot1962}. The fluid motion of the slow wave at low frequencies is diffusive and governed by Darcy's law, whereas that at high frequencies is propagatory and controlled by the tortuosity \cite{Johnson1987}. The fluid flow due to the diffusive slow wave or the effect of the static permeability can be measured at a borehole because the borehole fluid communicates with the pore fluid in the formation at the borehole wall. Therefore, understanding the physical mechanisms behind the dynamic interaction between the porous formation and the borehole fluid is crucial in interpreting data to estimate the hydraulic properties. 

In order to explore this dynamic interaction, extensive research has been conducted in the context of full-waveform acoustic logging using Stoneley (tube) waves \cite<e.g.,>[among many others]{Biot1952, Cheng1987, Chang1988, Fan2013, Sidler2014GJI}. The Stoneley wave is an axially symmetric wave having a fundamental mode in the borehole that exists from the zero frequency \cite{White1983}. At low frequencies, the Stoneley wave propagates as a piston-like compression of the borehole fluid, which is often called a tube wave \cite{Endo2006}. The tube wave is known to be dominant in data at low frequencies \cite{Tang1993}. In acoustic logging using the tube waves, a pressure pulse is generated in the borehole at relatively low frequencies ($\sim$2 kHz), where the interaction between the tube wave and the diffusive slow wave is prominent. The tube waves propagate along the borehole. Tube-wave amplitude is attenuated at the intersection of the borehole fluid and the porous formation, where a part of the energy is carried away by the slow wave \cite{Tang1991}. Attenuation, dispersion, and reflection of tube waves have been analyzed in the past in order to infer hydraulic permeability \cite<e.g.,>[]{Tang1993, Tang1996, Endo2006}. The high sensitivity of the tube wave to the surrounding formation's permeability has also been exploited in monitoring a borehole during fluid production \cite{Bakulin2008}.

Vertical seismic profiling (VSP) also measures the dynamic interaction between porous formations and the borehole fluid. In contrast to the full-waveform acoustic logging, VSP measures the borehole response of the low-frequency ($\sim$maximum a few hundred hertz) wavefield generated by a seismic source located at the surface. VSP is often performed in order to estimate the seismic velocity structures around a borehole using elastic body waves, where the tube waves are considered as noise \cite<e.g.,>[]{Hardage1981}. However, on many other occasions, exploiting the fact that the tube waves have large amplitudes and often dominate the fluid pressure in the borehole, hydrophone VSP data have been used in order to estimate in-situ permeability \cite{Huang1984,Beydoun1985,Hardin1987,Li1994,Kiguchi2001}. In VSP measurements, the amount of fluid flow due to dynamic poroelasticity is larger than the fluid flow in acoustic logging if one assumes a fixed pore-pressure gradient at the borehole wall. This difference is caused by greater dynamic permeability at lower frequencies \cite{Tang1991}. Furthermore, in the case of VSP, the energy of the tube waves increases at the presence of a porous formation, i.e., large-amplitude tube waves are generated due to conversion from elastic waves. This energy increase contrasts with the acoustic logging where the total energy of tube waves decreases due to diffusion of the borehole pressure at the intersection with the porous formation. Because of these differences, one can argue that hydrophone VSP analysis is more sensitive to hydraulically active zones. Note that hydrophone VSP has been commended for many years to achieve a significant reduction of the data acquisition cost \cite<e.g.,>[]{Marzetta1988,Milligan1997,Greenwood2012}. More recently, \citeA{Greenwood2019Tec} show that tube waves observed using a slotted PVC casing are similar to those detected using an open hole. These studies suggest the possibility of efficient permeability monitoring using VSP, without the long-term maintenance of an open borehole.

\begin{figure}
\centering
 \noindent\includegraphics{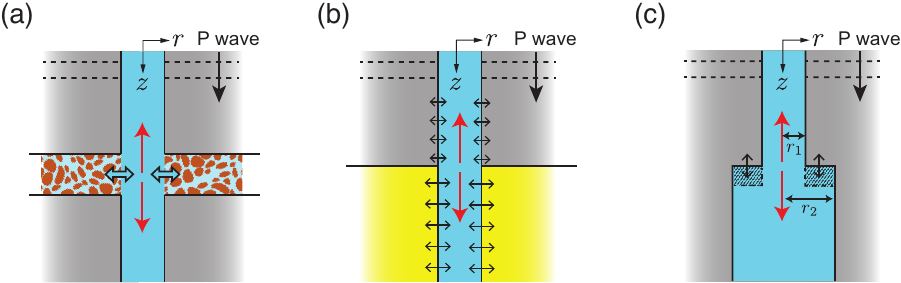}
\caption{Illustration of three different mechanisms that generate Stoneley (tube) waves (red arrows) due to an incident P wave. The cylindrical coordinate system ($r$-$z$) is considered. (a) Fluid infiltration (blue arrows) from the porous formation. (b) The difference in the degree of radial strain (black arrows) at the borehole wall at the elastic layer boundary. (c) The local change of fluid volume (shaded area) at the borehole irregularity (a step-like change in the borehole radius).  }
\label{fig:intro_3M}
\end{figure}

In VSP measurements, the physical mechanisms that affect primarily the fluid pressure in a borehole are different from those in acoustic logging. The major difference is the presence of externally propagating elastic waves due to a source at the surface. Three different mechanisms underlying the pressure perturbation due to the disturbance of P waves have been discussed in the past \cite<e.g.,>[]{White1983,Beydoun1985,Peng1996}, which is summarized in Figure \ref{fig:intro_3M}. In a pioneering work, \citeA{Beydoun1985} developed a model to predict the amplitude of tube waves generated at open fractures due to fracture deformation by elastic waves. The model considers the fluid volume at the intersection between the open fracture and the borehole (schematically shown in Figure \ref{fig:intro_3M}a). Several open-fracture or porous-formation models, that relate the fluid volume and the amplitude of the tube wave, have subsequently been developed \cite<e.g.,>[]{Hardin1987, Li1994, Ionov2007, Bakku2013, Minato2017SEG, Minato2017JGR}. The tube waves in VSP have been investigated also in the context of the borehole coupling theory for elastic waves \cite<e.g.,>[]{White1953,White1983,Schoenberg1986}. \citeA{Peng1994PhD} showed the relation between the fluid pressure and the borehole squeeze strain (azimuthally averaged radial strain at the borehole wall) due to the elastic waves. This relation has been used in order to calculate the complete borehole pressure response (i.e., including the response of elastic waves and tube waves) in VSP or crosshole configuration \cite{Kurkjian1994,Peng1996}. In this case, the difference in the borehole squeeze strain along depth, which is due to the differences in the elastic properties of the formation, produces the tube wave at the boundary (Figure \ref{fig:intro_3M}b). Finally, tube waves are generated at the boundary where the borehole radius abruptly changes \cite{Hardage1981,White1988}, which is caused by the piston-like movement of an annular ledge at the boundary (Figure \ref{fig:intro_3M}c). Note that, although we focus on tube waves caused by P waves in this study, it is known that tube waves are generated also due to S waves \cite{Toksoz1992} and the interaction between the surface wave and the top of the borehole \cite{Hardage1981}. In VSP at a relatively deep (a few hundred meters) borehole, such tube waves arrive much later than P waves. In this case, it is trivial to isolate tube waves excited by P waves.

\begin{figure}
\centering
 \noindent\includegraphics{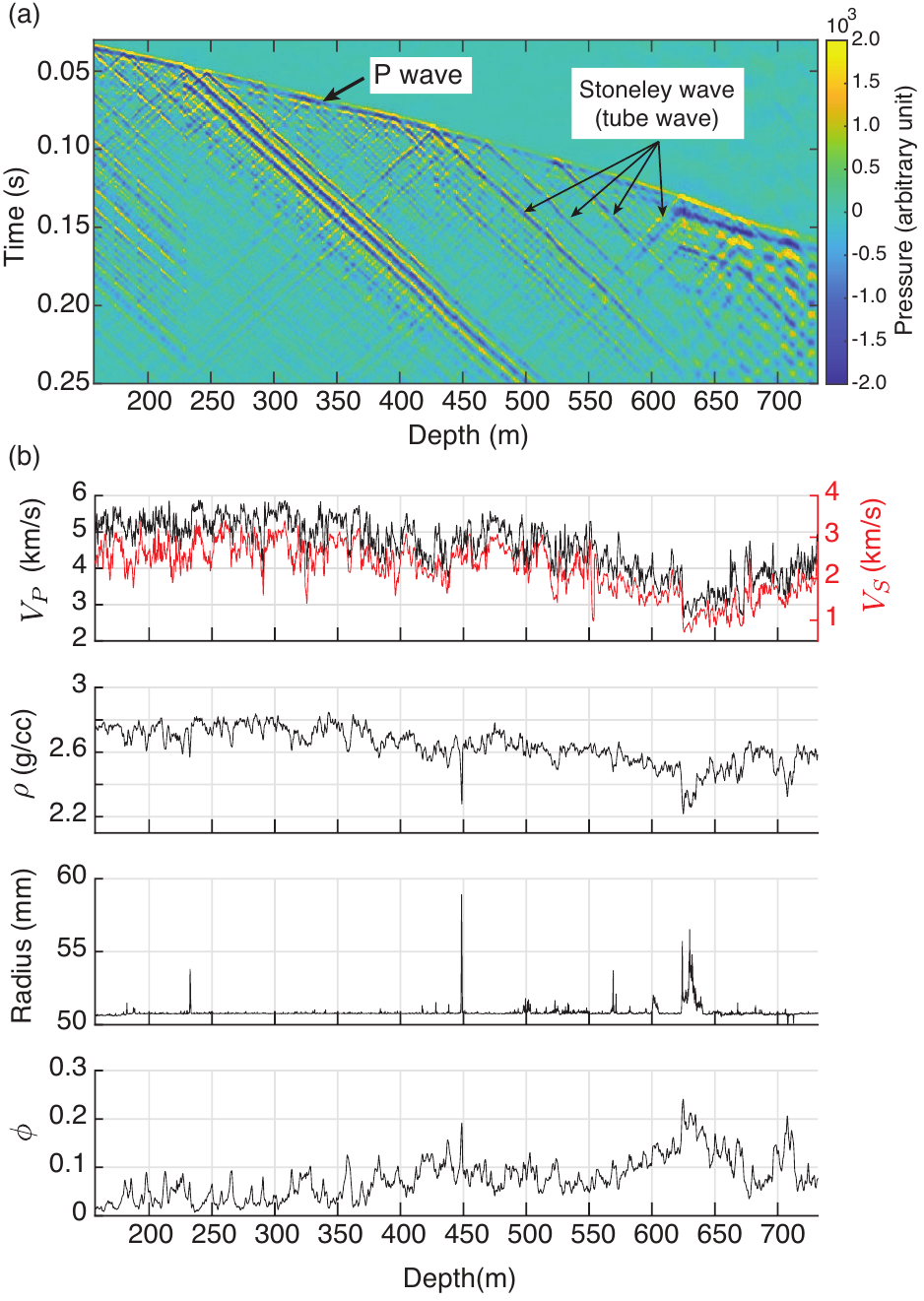}
\caption{(a) Fluid pressure response at a borehole located in a fault zone due to incident P waves from a pressure source at the surface. (b) Downhole log data: P-wave velocity ($V_P$), S-wave velocity ($V_S$), density ($\rho$), borehole radius, and porosity ($\phi$). }
\label{fig:intro_data}
\end{figure}

It is clear that all the mechanisms mentioned above produce tube waves in a borehole. However, the role of each mechanism in a quantitative interpretation of observed tube waves is poorly understood and, to our knowledge, has rarely been discussed. Correlations have been shown between the depth of tube-wave generation and the known locations of the geological features (permeable structures, impedance boundaries, borehole irregularities) in log data \cite<e.g.,>[]{Beydoun1985,Li1994,Kiguchi2001,Evans2005,Greenwood2019Tec}. But only looking at these correlations does not allow quantitative interpretation of data at a fault zone, where all of the above mechanisms can potentially contribute. Figure \ref{fig:intro_data}(a) shows field hydrophone VSP data measured at an active fault zone in Japan \cite{Kiguchi2001}. The survey depth interval contains open fractures and fault-related permeable structures, e.g., cataclasites and fault breccias. These data unequivocally illustrate that tube waves are generated at discrete depths due to an incident P wave. Some of these depths correlate well with the location of the permeable structures \cite{Kiguchi2001}. However, the downhole log data (Figure \ref{fig:intro_data}b) show a considerable variation in seismic velocities, density, borehole radius, and porosity along the borehole. Due to a lack of forward modeling approaches and analytical solutions that predict pressure amplitudes of the borehole fluid taking into account all three possible mechanisms, it is still unknown how these heterogeneities have played a role in the observed data. 

In this study, we develop the theory that allows calculation of the borehole pressure response when a P wave is incident on an irregular borehole surrounded by layered poroelastic media, taking into account simultaneously all three possible mechanisms (Figure \ref{fig:intro_3M}). \citeA{Ionov1996} notably contributed to the study of tube-wave propagation. They developed a theory that introduces fluid flow due to a porous formation (Figure \ref{fig:intro_3M}a) into the borehole coupling theory (Figure \ref{fig:intro_3M}b). We supplement to this approach the effect of borehole irregularities (Figure \ref{fig:intro_3M}c), and extend the theory so that it is consistent with Biot poroelasticity at low frequencies. The new model simplifies the coupled acoustic-poroelastic problem into two subproblems: (1) the external elastic wave propagation in the absence of a borehole and (2) the fluid-pressure response due to inclusion of a cylindrical fluid column (borehole) based on the quasi-static/low-frequency approximation. As a result, this model can calculate complete waveforms including elastic waves and tube waves at low frequencies. Considering a normally incident plane P wave, we solve the system of equations using the propagator matrix method \cite{Aki2002_Ch7.2.2}. This leads to derivation of the closed-form analytical expressions for the amplitude of the generated tube waves due the three mechanisms illustrated in Figure \ref{fig:intro_3M}. The new formulations also enable calculating fast the pressure waveforms at complex structures, such as those shown in Figure \ref{fig:intro_data}, taking simultaneously into account all three possible mechanisms. 

In this study, furthermore, we verify the developed theory and the derived analytical solutions using finite-difference (FD) numerical solutions, explicitly taking into account wave propagation in a borehole embedded in poroelastic media. Despite the relatively simple configuration assumed in this study, the problem is complex, and solving it using the FD method is computationally demanding. This is because the problem is three-dimensional and multi-scale: while the seismic wavelength is in decameter-scale, the wave propagation is in hectometer-scale, and the observation takes place in a borehole with a radius which is in the centimeter-scale. Arguably, this complexity has so far hindered a detailed comparison among different tube-wave generation models and the complete numerical solutions. To our knowledge, such verification has so far been performed only for a borehole with a constant radius embedded in heterogeneous elastic media \cite{Peng1992SEG,Kurkjian1994}, i.e., only for the mechanism shown in Figure \ref{fig:intro_3M}(a). In this study, we mitigate the computational challenge through adopting a new approach that uses the FD method in an azimuthally symmetric, cylindrical coordinate system and with initial conditions that simulate plane P-wave incidence. 

We first present the governing equations and the propagator-matrix formulation to solve these equations considering multi-layered media. We then derive new, closed-form, analytical expressions for the amplitude of the tube waves incorporating each mechanism individually: the elastic-layer boundary (Figure \ref{fig:intro_3M}b), a porous layer sandwiched between two elastic layers (Figure \ref{fig:intro_3M}a), and a step-like change in the borehole radius (Figure \ref{fig:intro_3M}c). The relevance of the derived expressions in comparison with solutions obtained in earlier studies is discussed. Finally, we present the numerical modeling results for the complex structure at a fault zone (Figure \ref{fig:intro_data}) and discuss the role of different mechanisms in field data. 

\section{A simplified theory to calculate acoustic-poroelastic coupling at an irregular borehole due to an incident plane P wave}
\label{Sec:Theory}
\subsection{Motion of the borehole fluid surrounded by poroelastic media due to elastic wave propagation}
We start with the equations of \citeA{Ionov1996} in order to describe the low-frequency wave motion at a fluid-filled borehole surrounded by poroelastic media. The theory calculates the complete pressure response including elastic waves and tube waves. This earlier model includes tube waves generated at the porous formation where fluid infiltration occurs (Figure \ref{fig:intro_3M}a) and also tube waves generated at the boundary between two elastic media (Figure \ref{fig:intro_3M}b). However, it does not contain tube waves generated at the borehole irregularities (Figure \ref{fig:intro_3M}c): this will be introduced as an additional point source in the propagator matrix formulation in a later subsection. We reformulate the governing equations of \citeA{Ionov1996} to solve them by the propagator matrix method \cite{Aki2002_Ch7.2.2}. Furthermore, we introduce a new boundary condition such that the theory becomes consistent with Biot dynamic poroelasticity at low frequencies.

We consider a small-amplitude wave motion of a circular fluid-cylinder surrounded by poroelastic media (i.e., an open borehole). At low frequencies where the seismic wavelength is much larger than the borehole radius, pressure ($p$) and vertical particle velocity ($v_z$) of the borehole fluid satisfy the following partial differential equation (see \ref{Append:Ionov} for more details):
\begin{linenomath*}
 \begin{equation}
  \frac{\partial}{\partial z}
   \begin{pmatrix}
    p \\ 
    v_z
   \end{pmatrix}=
  i\omega\begin{pmatrix}
	  0 & \rho_f \\
	  K_{\rm eff}^{-1} & 0
	 \end{pmatrix}
         \begin{pmatrix}
	  p \\ 
	  v_z
	 \end{pmatrix}+
  i\omega\begin{pmatrix}
	  0 \\ 
	  \frac{2\sigma^{\rm ext}_{\rm eff}}{E}-\frac{2\phi}{K_f}p^{\rm ext}_{\rm por}\Phi\left(\sqrt{-i\omega t_f}\right)
	 \end{pmatrix},
	 \label{eq:PDE1}
 \end{equation}
\end{linenomath*}
where $\rho_f$ is the density of the fluid, $K_f$ the fluid bulk modulus, $K_{\rm eff}$ the effective bulk modulus of the borehole fluid, $E$ the Young's modulus of the elastic medium surrounding the borehole, $\phi$ the porosity of the porous formation, and $\Phi$ the function relevant to fluid infiltration at the boundary between the borehole and the porous formation (see \ref{Append:Ionov}). The effective external stress ($\sigma^{\rm ext}_{\rm eff}$) and the pore pressure away from the borehole ($p^{\rm ext}_{\rm por}$) characterize the effect due to the externally propagating elastic waves, which will be explained shortly. Equation \ref{eq:PDE1} is derived by averaging the continuity equation and the equation of motion of the borehole fluid over the borehole cross-section and relating the horizontal motion of the borehole wall to that due to elastic deformation and additional fluid infiltration \cite{Ionov1996}. Note that in the model of \citeA{Ionov1996}, the deformation of the elastic medium and the fluid infiltration from the porous formation are separately considered under the low-frequency approximation. Therefore, any elastic property (e.g., elastic moduli and velocities) is understood as that at the elastic limit or the Gassmann's low-frequency limit of the poroelastic moduli (\ref{Append:B}). In equation \ref{eq:PDE1}, $K_{\rm eff}$ is related to the formation and fluid properties as,
\begin{linenomath*}
 \begin{equation}
  K_{\rm eff}^{-1}=K_f^{-1}+\mu^{-1}+\frac{2\phi}{K_f}\Phi\left(\sqrt{-i\omega t_f}\right),  \label{eq:Keff}
 \end{equation}
\end{linenomath*}
where $\mu$ is the shear modulus of the surrounding formation. The phase velocity of the tube wave ($C_T$) is defined as $C_T=\sqrt{K_{\rm eff}/\rho_f}$. 

 The source term (the second term on the right-hand side of equation \ref{eq:PDE1}) characterizes the response of the borehole fluid due to an elastic wave propagating in the formation. In this study, we consider a normal incident plane P wave with the displacement potential in the form of $\phi_{\rm E}(z)=D_{\rm E}\exp(ik_pz)+U_{\rm E}\exp(-ik_pz)$, where the subscript ``E'' stands for elastic wave and $k_p=\omega/V_P$ (see also \ref{Append:phiE}). In this case, the source term associated with the effective external stress ($\sigma^{\rm ext}_{\rm eff}$) represents the elastic deformation of the wall due to the P wave \cite{Ionov1996}:
\begin{linenomath*}
\begin{align}
 \sigma^{\rm ext}_{\rm eff}(z)&=\sigma_{rr}+\sigma_{\theta\theta}-\nu\sigma_{zz} \\
                              &=-E\omega^2\left(\frac{1}{2V_S^2}-\frac{1}{V_P^2}\right)\phi_{\rm E}(z),
\label{eq:sext}
\end{align}
\end{linenomath*}
where $V_P$ and $V_S$ are P- and S-wave velocities of the formation. Note that $\sigma^{\rm ext}_{\rm eff}/E$ is known as the borehole squeeze strain in the borehole coupling theory \cite{Peng1996}. 

The remaining source term in equation \ref{eq:PDE1} or that associated with $p^{\rm ext}_{\rm por}$ represents the fluid flow at the borehole wall which intersects the porous formation. Due to the low-frequency approximation, the fluid-flow system is assumed to be governed by the viscous force (see Equation \ref{eq:PDE_por}). Furthermore, the elastic deformation (derived from Gassmann's elastic moduli) and the fluid flow are related through the boundary condition for the pore pressure ($p^{\rm ext}_{\rm por}$). This indicates that, although the theory discussed here assumes a low-frequency limit of Biot poroelasticity where the wavefield becomes purely elastic, the additional boundary condition at the intersection between the borehole fluid and the porous formation is responsible for the fluid flow in the model. The pore pressure away from the borehole ($p^{\rm ext}_{\rm por}$), which is generated by the external seismic wavefield, can be represented as (see \ref{Append:Ionov} for more details),
\begin{linenomath*}
\begin{align}
 p^{\rm ext}_{\rm por}(z)&=-\frac{1}{3}B\sigma_{ii} \nonumber \\
                         &=Bk_p^2 K\phi_{\rm E}(z),
\label{eq:pext}
\end{align}
\end{linenomath*}
where $\sigma_{ii}$ is the trace of the elastic stress tensor, $K$ is the bulk modulus of the elastic formation, and $B$ is the Skempton coefficient (see \ref{Append:B}). Note that in the original formulation in \citeA{Ionov1996}, the effect of the Skempton coefficient is not taken into account. We show in a later section that this new boundary condition is necessary so that the theory is consistent with the numerical results using Biot dynamic poroelasticity.

\subsection{Propagator matrix formulation to calculate the borehole response}
\begin{figure}
\centering
 \noindent\includegraphics{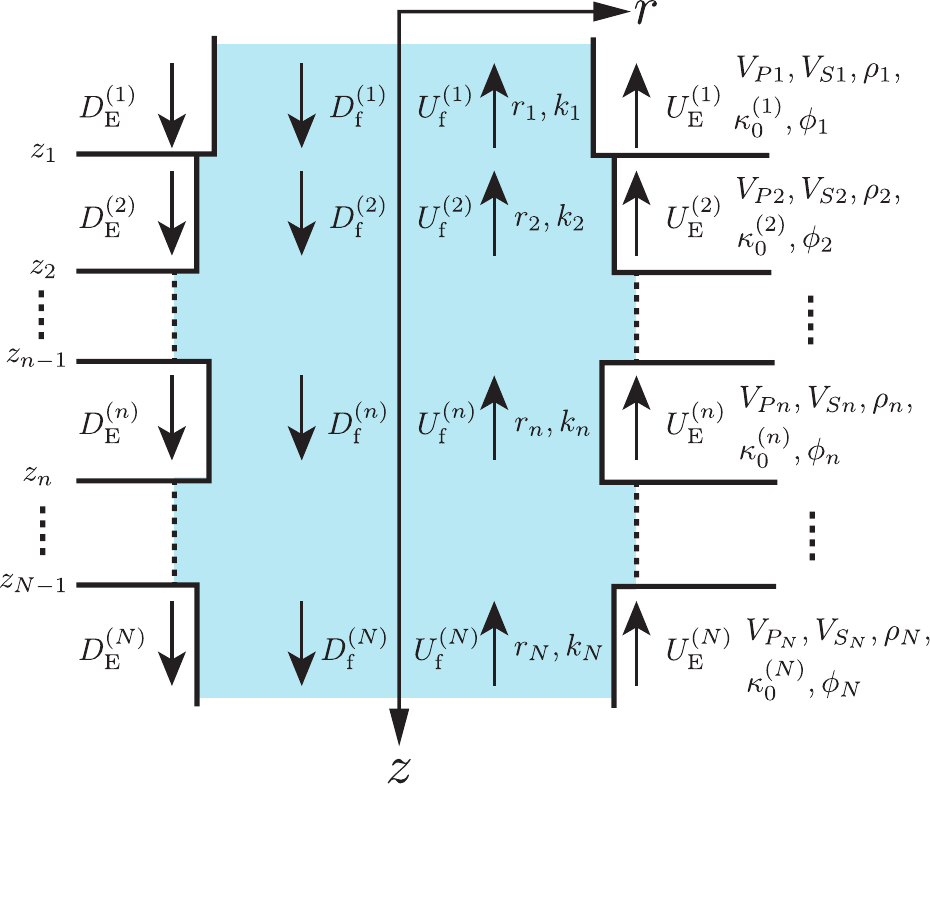}
\caption{An irregular borehole in a multi-layered poroelastic subsurface. The $N$-layers are described by their respective parameters and the borehole radius. The upgoing and downgoing potential amplitudes at each layer are indicated as $U^{(i)}_{\rm f}$ and $D^{(i)}_{\rm f}$ for the tube wave, and $U^{(i)}_{\rm E}$ and $D^{(i)}_{\rm E}$ for the elastic wave.}
\label{fig:Geom_General}
\end{figure}

Assuming a stack of $N$ horizontal layers ($N>2$), we define the potential function of the borehole fluid in the form of $\phi_{\rm f}(z)=D_{\rm f}\exp(ikz)+U_{\rm f}\exp(-ikz)$, where the subscript ``f'' stands for the fluid, and the tube-wave wavenumber $k=\omega/C_T$ (Figure \ref{fig:Geom_General}). We solve the governing equation (equation \ref{eq:PDE1}) for the potential amplitudes ($D_{\rm f}$ and $U_{\rm f}$) using the propagator matrix method in order to calculate the borehole response. Note that the pressure and the vertical velocity of the borehole fluid in the source-free region are defined as $p=\rho_f\omega^2\phi_{\rm f}$ and $v_z=-i\omega\partial \phi_{\rm f}/\partial z$, respectively. As explained earlier, the theory simplifies the problem into two subproblems (elastic wave propagation and borehole-fluid response). Consequently, the potential function of the P wave ($\phi_{\rm E}$) is presumed to have been obtained by another propagator matrix approach (see \ref{Append:phiE}).

At the {\it n}-th boundary ($z=z_n$), we specify the following boundary conditions:
\begin{linenomath*}
 \begin{align}
  p^{(n)}+\Delta p^{(n)}&=p^{(n+1)}, \label{eq:BC1} \\
  \pi r_n^2 \{v_z^{(n)}+\Delta v_z^{(n)}\}&=\pi r_{n+1}^2 v_z^{(n+1)}, \label{eq:BC2}
 \end{align}
\end{linenomath*}
where $p^{(n)}$ and $v_z^{(n)}$ are the pressure and the vertical velocity of the borehole fluid in the absence of the source, and $r_n$ is the borehole radius at the {\it n}-th layer (Figure \ref{fig:Geom_General}). These boundary conditions describe the continuity of the pressure and that of the flow volume across the boundary \cite<e.g.,>[]{Ionov1996,Tezuka1997}. The discontinuities $\Delta p^{(n)}$ and $\Delta v_z^{(n)}$ contain contributions due to the external elastic waves or the three mechanisms shown in Figure \ref{fig:intro_3M} (see \ref{Append:delta}). The details of the discontinuities are discussed in the following subsection. 

From the boundary conditions (equations \ref{eq:BC1} and \ref{eq:BC2}), the potential-amplitude vector at the {\it n}-th layer, i.e., $\mathbf{u}_n=(U_{\rm f}^{(n)},D_{\rm f}^{(n)})^{\rm T}$, is represented by the following relation:
\begin{linenomath*}
 \begin{equation}
 \mathbf{u}_n=\mathbf{M}_n(z_n)\mathbf{u}_{n+1}+\mathbf{S}_n(z_n).
  \label{eq:un}
 \end{equation}
\end{linenomath*}
 The matrix $\mathbf{M}_n$ is defined as
\begin{linenomath*}
 \begin{align}
 \mathbf{M}_n(z)&=\frac{1}{2 r_n^2k_n}
   \begin{pmatrix}
    a_1 e^{i(k_n-k_{n+1})z} && a_2 e^{i(k_n+k_{n+1})z} \\ 
    a_2 e^{-i(k_n+k_{n+1})z} && a_1 e^{-i(k_n-k_{n+1})z} 
   \end{pmatrix},
    \label{eq:Mn} \\
 a_1&=r_n^2k_n+r_{n+1}^2k_{n+1}, \\
 a_2&=r_n^2k_n-r_{n+1}^2k_{n+1},
 \end{align}
\end{linenomath*}
where $k_n$ is the tube-wave wavenumber at the {\it n}-th layer, and the source vector $\mathbf{S}_n$ is defined as
\begin{linenomath*}
 \begin{equation}
 \mathbf{S}_n(z)=\frac{1}{2 \rho_f \omega^2 k_n}
   \begin{pmatrix}
    \{\Delta v_z^{(n)}\rho_f \omega-k_n\Delta p^{(n)}\} e^{i k_n z}  \\ 
    -\{\Delta v_z^{(n)}\rho_f \omega+k_n\Delta p^{(n)}\} e^{-i k_n z} 
   \end{pmatrix}.
    \label{eq:Sn}
 \end{equation}
\end{linenomath*}
Successively applying equation \ref{eq:un}, we obtain the following relation:
\begin{linenomath*}
 \begin{align}
 \mathbf{u}_1&=\prod_{i=1}^{N-1}\mathbf{M}_i(z_i)\mathbf{u}_{N}+\left[\sum_{j=2}^{N-1}\prod_{i=2}^{j}\mathbf{M}_{i-1}(z_{i-1})\mathbf{S}_{j}+\mathbf{S}_1\right] \nonumber \\
  &=\mathbf{M}_T\mathbf{u}_N+\mathbf{S}_T.
  \label{eq:unu1} 
 \end{align}
\end{linenomath*}
Equation \ref{eq:unu1} assumes $N>2$, and the matrix $\mathbf{S}_T$ contains contributions of the source terms accumulated from the first boundary ($z=z_1$). In the case of a single boundary ($N=2$), a particular treatment is required (see \ref{Append:homo} and Section \ref{Sec:Elastic}).

Once we solve equation \ref{eq:unu1} with appropriate radiation conditions (see Section \ref{Sec:Elastic} and \ref{Sec:TPL}), the potential amplitudes at all layers can be calculated using equation \ref{eq:un}. Finally, the borehole response, i.e., the pressure and the vertical velocity of the borehole fluid including the source term, can be obtained at the {\it n}-th layer as
\begin{linenomath*}
 \begin{align}
  p(z)&=\rho_f\omega^2\left(D^{(n)}_{\rm f} e^{ik_nz}+U^{(n)}_{\rm f} e^{-ik_nz}\right)+\Delta p^{(n)}(z), \label{eq:p} \\
 v_z(z)&=k_n\omega\left(D^{(n)}_{\rm f} e^{ik_nz}-U^{(n)}_{\rm f} e^{-ik_nz}\right)+\Delta v_z^{(n)}(z), \label{eq:vz} 
 \end{align}
\end{linenomath*}
where $z_{n-1} \le z \le z_n$, and $2\le n \le N-1$.

\subsection{Discontinuities due to continuous and point sources}
The discontinuities $\Delta p$ and $\Delta v_z$ in the boundary conditions (equations \ref{eq:BC1} and \ref{eq:BC2}) and those in the final borehole response (equations \ref{eq:p} and \ref{eq:vz}) represent the simultaneous effects of the three different mechanisms (Figure \ref{fig:intro_3M}). They are defined as
\begin{linenomath*}
 \begin{align}
 \Delta p^{(n)}(z)&=\Delta p_{\rm E}^{(n)}(z_{n-1},z)+\Delta p_{\rm ft}^{(n)}(z_{n-1},z),   \label{eq:delta_p} \\
 \Delta v_z^{(n)}(z)&=\Delta v_{\rm E}^{(n)}(z_{n-1},z)+\Delta v_{\rm ft}^{(n)}(z_{n-1},z)+\Delta v_{\rm q}^{(n)}\delta(z-z_n),
  \label{eq:delta_v}
 \end{align}
\end{linenomath*}
where the subscript ``E'' indicates the contribution due to elastic deformation (Figure \ref{fig:intro_3M}b), the subscript ``ft'' the contribution due to fluid infiltration (Figure \ref{fig:intro_3M}a), and the subscript ``q'' the contribution due to borehole irregularity (Figure \ref{fig:intro_3M}c). As we will see in the following, all these contributions are described by the potential amplitudes of the P wave.

The discontinuities due to the elastic deformation and the fluid infiltration (i.e., ``E'' and ``ft'') are calculated from the source located continuously over depth within each layer (equation \ref{eq:PDE1}). They are obtained using the propagator matrix formulation (\ref{Append:delta}) as,  
\begin{linenomath*}
 \begin{align}
 \Delta p^{(n)}_{\rm E}(z_{n-1},z)&=-i\omega\rho_f C_T k_p A_P \{D^{(n)}_{\rm E} I_1(z_{n-1},z)+U^{(n)}_{\rm E} I_2(z_{n-1},z)\}   \label{eq:delta_pE}\\
 \Delta v^{(n)}_{\rm E}(z_{n-1},z)&=-i\omega k_p A_P \{D^{(n)}_{\rm E} I_3(z_{n-1},z)+U^{(n)}_{\rm E} I_4(z_{n-1},z)\}
  \label{eq:delta_vE}, 
 \end{align}
\end{linenomath*}
and
\begin{linenomath*}
 \begin{align}
 \Delta p^{(n)}_{\rm ft}(z_{n-1},z)&=-i\omega  \rho_f C_T\frac{\phi\Phi K B}{K_f} k_p^2\{D^{(n)}_{\rm E} I_1(z_{n-1},z)+U^{(n)}_{\rm E} I_2(z_{n-1},z)\}   \label{eq:delta_pft} \\
 \Delta v^{(n)}_{\rm ft}(z_{n-1},z)&=-i\omega \frac{\phi\Phi K B}{K_f} k_p^2\{D^{(n)}_{\rm E} I_3(z_{n-1},z)+U^{(n)}_{\rm E} I_4(z_{n-1},z)\},
  \label{eq:delta_vft}
 \end{align}
\end{linenomath*}
where all material properties correspond to those at the {\it n}-th layer. The functions $I_1$, $I_2$, $I_3$, and $I_4$ characterize the interferences of the waves propagating with P-wave and tube-wave velocities within the layer (\ref{Append:delta}). The factor $A_P$ in equations \ref{eq:delta_pE} and \ref{eq:delta_vE} is the one for the elementary pressure pulse in the borehole \cite{Peng1992SEG}:
\begin{linenomath*}
 \begin{equation}
  A_P=\omega^2k_p^{-1}\left(\frac{1}{2V_S^2}-\frac{1}{V_P^2}\right). \label{eq:AP}
 \end{equation}
\end{linenomath*}

The discontinuity due to borehole irregularity ($\Delta v_{\rm q}$ in equation \ref{eq:delta_v}) is modeled as a point injection-rate source. Here we consider the step-like change of the borehole radius at a boundary (see Figure \ref{fig:intro_3M}c). In this case, the piston-like action of an annular ledge at the boundary produces a change in the fluid volume locally \cite<shaded area in Figure \ref{fig:intro_3M}c;>[]{Kurkjian1994}:
\begin{linenomath*}
 \begin{equation}
  \Delta V=\pi(r^2_{n+1}-r^2_n)v_z^E, \label{eq:dV}
 \end{equation}
\end{linenomath*}
where $\Delta V$ is the rate of the fluid-volume change ($\rm m^3/s$) due to the vertical particle velocity of the external elastic wave $v_z^E$ at the boundary ($z=z_n$). The discontinuity $\Delta v_{\rm q}$ can be written in terms of the injection-rate source in the unit volume as,
\begin{linenomath*}
 \begin{align}
  \Delta v_{\rm q}^{(n)}&=\frac{\Delta V}{\pi r_n^2} \nonumber \\
            &=\frac{r^2_{n+1}-r^2_n}{r^2_n}k_p\omega\left(D^{(n)}_E e^{i k_p z_n}-U^{(n)}_E e^{-i k_p z_n}\right),
 \end{align}
\end{linenomath*}
where we use the relation $v_z^E=-i\omega\partial \phi_{\rm E}/\partial z$. Note that the rate of the fluid-volume change (equation \ref{eq:dV}) assumes the boundary between the borehole fluid and the elastic medium \cite{Kurkjian1994}. At the boundary between the borehole fluid and a poroelastic medium, the same equation can be applied using the low-frequency elastic moduli (\ref{Append:B}). In this case, however, the fluid entering vertically from the poroelastic medium to the borehole is ignored. The effects of the vertical flow at the boundary depend primarily on the length of the annular ledge at the radius change. Using finite-difference modeling and considering a realistic borehole radius (i.e., $r$=50--100 mm), we found that such vertical flow is negligibly small compared to the elastic deformation ($v_z^E$). 

\section{Amplitude of the generated tube wave due to each mechanism}
\label{Sec:Simple_Config}

The theory developed above enables obtaining the closed-form expressions of the analytical amplitude of the generated tube waves due to normally incident plane P waves. The theory can include simultaneous effects due to the three mechanisms shown in Figure \ref{fig:intro_3M}. In this section, however, we derive the analytical amplitude for each mechanism considering the simple yet practically important configurations: the elastic layer boundary (Section \ref{Sec:Elastic}), the thin porous layer (Section \ref{Sec:TPL}), and the step-like change in borehole radius (Section \ref{Sec:Caliper}). We then verify using finite-difference (FD) numerical solutions, explicitly considering wave propagation at a cylindrical inclusion of acoustic fluid (borehole) embedded in 3D poroelastic media. We also discuss the derived expressions and their relevance to known solutions. 

\subsection{Elastic layer boundary}
\label{Sec:Elastic}
\subsubsection{Analytical expression of generated tube-wave amplitudes}
\begin{figure}
\centering
 \noindent\includegraphics{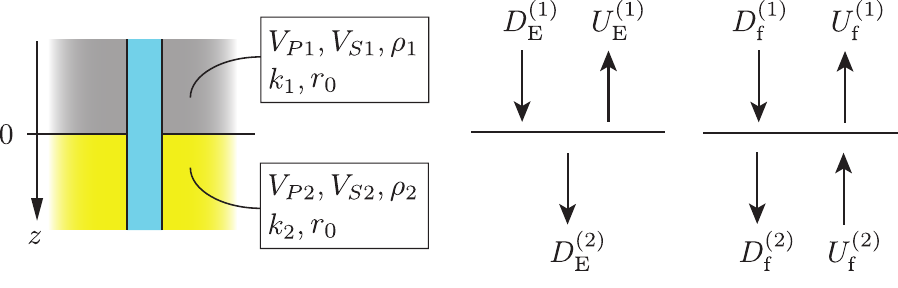}
\caption{A borehole with a constant radius ($r_0$) in two elastic half-spaces. The elastic-wave potential amplitudes contain the incident ($D_{\rm E}^{(1)}$), reflected ($U_{\rm E}^{(1)}$), and transmitted ($D_{\rm E}^{(2)}$) waves. The tube-wave potential amplitudes contain the upgoing and the downgoing waves at each layer ($U^{(i)}_{\rm f}$ and $D^{(i)}_{\rm f}$).}
\label{fig:Geom_Elastic}
\end{figure}

We first investigate a borehole located in two elastic half-spaces (Figure \ref{fig:Geom_Elastic}). When the half-spaces have different elastic properties, a tube wave is generated at their boundary due to an incident elastic wave. Although derivation of closed-form expressions in this scenario was attempted in the past \cite{Peng1992SEG}, we illustrate that our new solutions are consistent with the FD modeling results.

In the case of a single boundary ($N=2$) and constant borehole radius, the source term in equation \ref{eq:unu1} vanishes (see \ref{Append:homo}). Consequently, we obtain the following relation,
\begin{linenomath*}
 \begin{equation}
    \begin{pmatrix}
    U^{(1)}_{\rm f} \\
    D^{(1)}_{\rm f}
   \end{pmatrix}=\mathbf{M}_1(0)
    \begin{pmatrix}
    U^{(2)}_{\rm f} \\
    D^{(2)}_{\rm f}
   \end{pmatrix},
  \label{eq:un_2L_hetero}
 \end{equation}
\end{linenomath*}
where $\mathbf{M}_1\ne\mathbf{I}$ because $k_2 \ne k_1$ in equation \ref{eq:Mn}, and we assume $z_1=0$ for brevity. For homogeneous media (i.e., identical layer properties at upper and lower half-spaces), both upgoing and downgoing fluid-potential amplitudes are non-zero. In this case, the interaction among tube waves propagating in different directions and the continuous source due to the elastic deformation produces a single wave mode having the P-wave velocity (see \ref{Append:homo}). In the two-layer model considered in this subsection, however, an additional tube wave is generated at the boundary. Consequently, the fluid-potential amplitudes will deviate from those of homogeneous media to accommodate the additional upgoing tube wave in the upper half-space ($U^{(1)}_{\rm f}$) and the downgoing tube wave in the lower half-space ($D^{(2)}_{\rm f}$). Therefore, equation \ref{eq:un_2L_hetero} is solved for unknown $U^{(1)}_{\rm f}$ and $D^{(2)}_{\rm f}$, using known values of $D^{(1)}_{\rm f}$ and $U^{(2)}_{\rm f}$. The latter two amplitudes can be written based on the solutions for homogeneous media (\ref{Append:homo}) as,  
\begin{align}
     D_{\rm f}^{(1)} &= \frac{k_{p_1}}{k_1\left(k_1 - k_{p_1}\right)}A_P^{(1)} D^{(1)}_{\rm E} +\frac{k_{p_1}}{k_1\left(k_1+k_{p_1}\right)}A_P^{(1)} U^{(1)}_{\rm E},   \label{eq:Df1_2L} \\ 
    U_{\rm f}^{(2)} &= \frac{k_{p_2}}{k_2\left(k_2+k_{p_2}\right)}A_P^{(2)}D_{\rm E}^{(2)}. \label{eq:Uf2_2L}  
\end{align}
Equation \ref{eq:Df1_2L} contains the additional term associated with $U^{(1)}_{\rm E}$ compared to the amplitude in a homogeneous medium (equation \ref{eq:Df1}) because of the presence of the reflected elastic wave in the upper half-space (Figure \ref{fig:Geom_Elastic}). The factors $A_P^{(1)}$ and $A_P^{(2)}$ are calculated by equation \ref{eq:AP} with the material properties at the upper and lower half-spaces, respectively. 

Using equations \ref{eq:un_2L_hetero}, \ref{eq:Df1_2L}, and \ref{eq:Uf2_2L}, we can obtain all potential amplitudes ($U^{(1)}_{\rm f}$, $D^{(1)}_{\rm f}$,$U^{(2)}_{\rm f}$, $D^{(2)}_{\rm f}$). Similar to the case of homogeneous media (\ref{Append:homo}), the analytical amplitudes of the borehole response can be obtained by calculating pressure at $z=\pm Z$ where $Z>0$, using equation \ref{eq:p}. In the upper half-space ($z=-Z$), we obtain
\begin{align}
     p(-Z) =&-\rho_f C_T^{(1)} \omega\frac{2 k_1 k_{p_1}}{k_{p_1}^2-k_1^2}A^{(1)}_P D_{\rm E}^{(1)} e^{i k_{p_1}(-Z)} \nonumber \\
            &-\rho_f C_T^{(1)} \omega\frac{2 k_1 k_{p_1}}{k_{p_1}^2-k_1^2}A^{(1)}_P R_{\rm E} D_{\rm E}^{(1)} e^{-i k_{p_1}(-Z)}  \nonumber \\
            &+ A_{\uparrow} e^{-i k_{1}(-Z)}. \label{eq:p_2L_up}
\end{align}
From the depth-dependent phase delay of each term in equation \ref{eq:p_2L_up}, one can identify that the first term is the downgoing direct P wave, the second term the upgoing reflected P wave, and the third term the generated tube wave propagating upward. The amplitude of the generated tube wave ($A_{\uparrow}$) is written as,
\begin{align}
            A_{\uparrow}=&\rho_f C_T^{(1)} \omega\frac{2 k_1 k_{p_1}}{k_{p_1}^2-k_1^2}A^{(1)}_P \left[\frac{k_2+k_{p_1}}{k_1+k_2}R_{\rm E}+\frac{k_2-k_{p_1}}{k_1+k_2}\right]D_{\rm E}^{(1)} \nonumber \\
              &+\rho_f C_T^{(2)} \omega\frac{2 k_2 k_{p_2}}{(k_1+k_2)(k_2+k_{p_2})}A^{(2)}_P T_{\rm E}D_{\rm E}^{(1)}
, \label{eq:Aup}
\end{align}
where the symbol $\uparrow$ indicates an upward propagating tube wave, and $R_{\rm E}$ and $T_{\rm E}$ are the P-wave reflection and transmission coefficients defined by the potential amplitudes (\ref{Append:phiE}). Note that the analytical tube-wave amplitude derived in this study (equation \ref{eq:Aup}) differs from the earlier study \cite{Peng1992SEG}. Similarly, the pressure response at the lower half-space ($z=+Z$) can be written as,
\begin{align}
     p(+Z) =& -\rho_f C_T^{(2)} \omega\frac{2 k_2 k_{p_2}}{k_{p_2}^2-k_2^2}A^{(2)}_P T_{\rm E} D_{\rm E}^{(1)} e^{i k_{p_1}Z} \nonumber \\
            & + A_{\downarrow} e^{i k_{1}Z}, \label{eq:p_2L_low}
\end{align}
where the first term shows the transmitted P wave and the second term the generated tube wave. The amplitude of the downgoing tube wave is expressed as,
\begin{align}
            A_{\downarrow}=&\rho_f C_T^{(2)} \omega\frac{2 k_2 k_{p_2}}{k_{p_2}^2-k_2^2}A^{(2)}_P \frac{k_1+k_{p_2}}{k_1+k_2}T_{\rm E}D_{\rm E}^{(1)} \nonumber \\
              &+\rho_f C_T^{(1)} \omega\frac{2 k_1 k_{p_1}}{k_1+k_2}A^{(1)}_P \left[\frac{R_{\rm E}}{k_1+k_{p_1}}+\frac{1}{k_1-k_{p_1}}\right]D_{\rm E}^{(1)}.
 \label{eq:Adown}
\end{align}

\begin{figure}
\centering
 \noindent\includegraphics{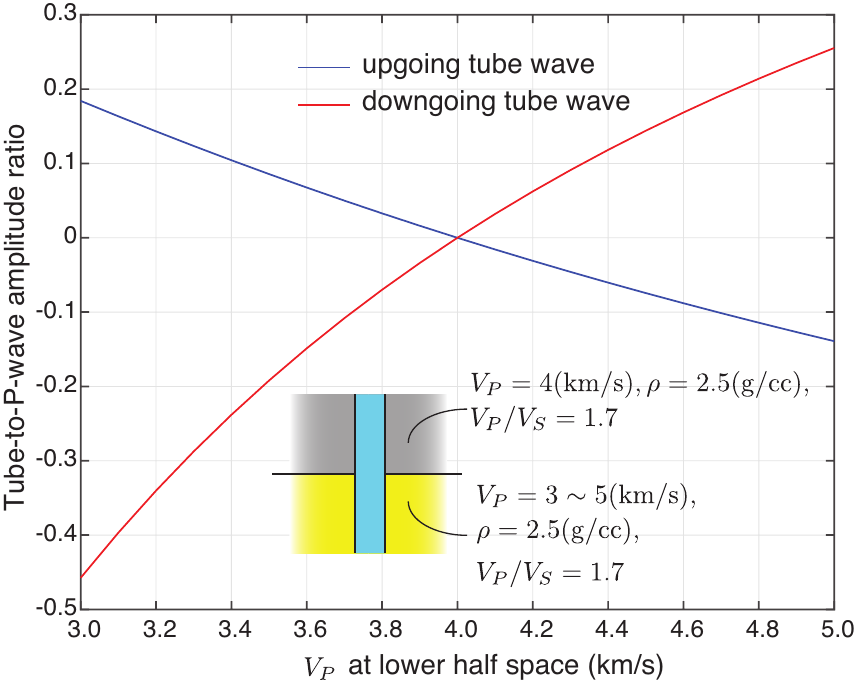}
\caption{Pressure amplitude of the generated tube wave (equations \ref{eq:Aup} and \ref{eq:Adown}) normalized by the amplitude of the incident P wave (the first term in equation \ref{eq:p_2L_up}) when the borehole is embedded in two elastic half-spaces. A fixed seismic velocity at the upper half-space and various seismic velocities at the lower space is used.}
\label{fig:Elastic_Ratio}
\end{figure}

\citeA{Peng1992SEG} first identified that the tube waves generated at the elastic impedance boundary and propagating in upward and downward directions have opposite polarities based on their approximate solutions. Here, we examine the polarity difference using the new analytical solutions (equations \ref{eq:p_2L_up} to \ref{eq:Adown}). Figure \ref{fig:Elastic_Ratio} shows the upgoing and downgoing tube wave amplitudes for various values for the elastic impedance contrasts. We evaluate the amplitude ratio between the incident P wave (the first term in equation \ref{eq:p_2L_up}) and the generated tube wave (equations \ref{eq:Aup} and \ref{eq:Adown}). From equations \ref{eq:AP}, \ref{eq:p_2L_up}, \ref{eq:Aup}, \ref{eq:Adown}, and \ref{eq:RTcoef_E}, one can see that the amplitude ratio is independent of the frequency. In Figure \ref{fig:Elastic_Ratio}, we consider the fixed layer properties in the upper half-space ($V_P=4$ km/s, $V_P/V_S=1.7$, $\rho=2.5$ g/cc) and the variable properties in the lower half-space ($V_P=$3--5 km/s, $V_P/V_S=1.7$, $\rho=2.5$ g/cc). The polarity difference is evident in upgoing and downgoing tube waves. The sign of the amplitudes is reversed when the lower half-space has higher seismic velocities than in the upper half-space (Figure \ref{fig:Elastic_Ratio}). 

\subsubsection{Verification through finite-difference modeling}

We verify the analytical solutions obtained in the previous subsection using numerical solutions. To this end, we use a staggered-grid finite-difference (FD) method in the cylindrical coordinate system \cite<e.g.,>[]{Randall1991, Mittet1996, Guan2011, Ou2019GJI}. The FD modeling solves Biot dynamic poroelasticity. In acoustic (borehole fluid) and elastic regions, we use the limiting values for the poroelastic properties (\ref{Append:Biot}). Note that, as we explained earlier, the verification of the approximate solutions based on the borehole coupling theory using the FD solution was performed in the past for a qualitative comparison, assuming a complex structure \cite{Kurkjian1994} or that assuming a high frequency (2 kHz) wavefield \cite{Peng1992SEG}. In this study, we quantitatively compare the amplitudes of each wave mode at low frequencies ($\sim$200 Hz) that represent field VSP measurements. 

\begin{figure}
\centering
 \noindent\includegraphics{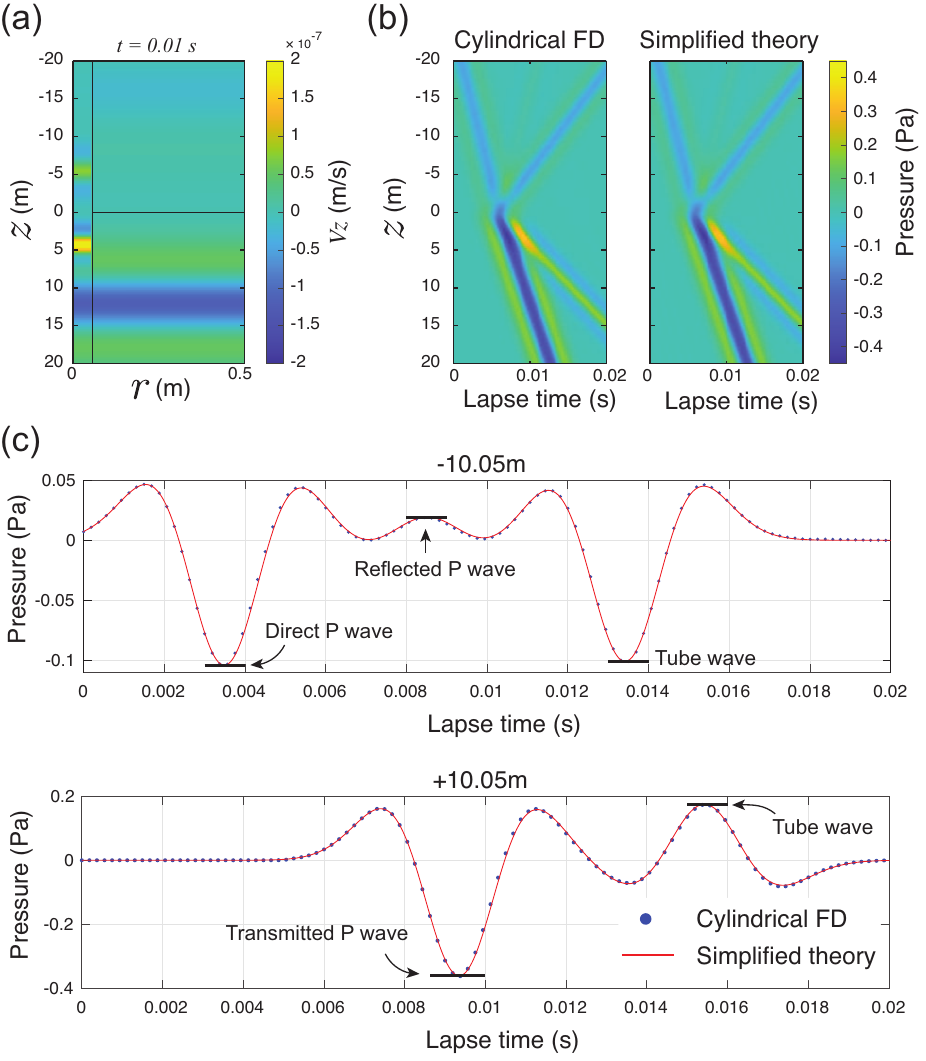}
\caption{Verification of the theory developed in this study using FD modeling for a borehole which is located in the two elastic half-spaces (Figure \ref{fig:Geom_Elastic}). (a) The snapshot of the wavefield (vertical particle velocity) calculated by cylindrical FD. The black vertical line indicates the location of the borehole wall, and the black horizontal line indicates that of the elastic impedance boundary. (b) Comparison of the pressure waveforms in the borehole calculated by FD and the theory developed in this study. (c) Same as (b) but for a receiver located at $\pm 10.05$ m from the elastic impedance boundary. The red lines are calculated waveforms using the new theory and the propagator matrix method (equations \ref{eq:unu1} and \ref{eq:p}); the black lines are analytical amplitudes derived from this theory (equations \ref{eq:p_2L_up} to \ref{eq:Adown}). }
\label{fig:Elastic_Layer}
\end{figure}

We assume the properties of the upper elastic half-space as $V_{P1}=4$ km/s, $V_{S1}=3$ km/s, $\rho_1=2.5$ g/cc, and those of the lower elastic half-space as $V_{P2}=3$ km/s, $V_{S2}=1$ km/s, $\rho_2=2.3$ g/cc. Figure \ref{fig:Elastic_Layer}(a) shows a snapshot of the FD modeling results or the distribution of the vertical particle velocity at $t=0.01$ s. In the FD modeling, we consider a borehole with a radius of 0.055 m. The grid spacing in FD modeling is 0.01 m in the radial direction and 0.1 m in the vertical direction. Unit-amplitude Ricker wavelet of 200 Hz center frequency is assumed for the vertical stress corresponding to the incident plane P wave. The initial condition of the FD modeling is set up such that the normal-incident plane P wave starts to propagate downward from 24 m above the elastic impedance boundary (\ref{Append:Biot}). The P wave propagates through the elastic impedance boundary at $z=0$ m, and consequently, the tube waves are generated and propagate 5 m from the boundary (Figure \ref{fig:Elastic_Layer}a). Figures \ref{fig:Elastic_Layer}(b) and (c) compare the borehole fluid pressure response using the FD modeling and that using the simplified theory developed in this study. The pressure values at $\pm 10.05$ m from the boundary show excellent agreement between the simplified theory and the complete numerical solution (Figure \ref{fig:Elastic_Layer}c). The analytical estimation (equations \ref{eq:p_2L_up} to \ref{eq:Adown}) correctly calculates the amplitudes for all wave modes (solid black lines in Figure \ref{fig:Elastic_Layer}c). 

\subsection{Thin porous layer}
\label{Sec:TPL}
\begin{figure}
\centering
 \noindent\includegraphics{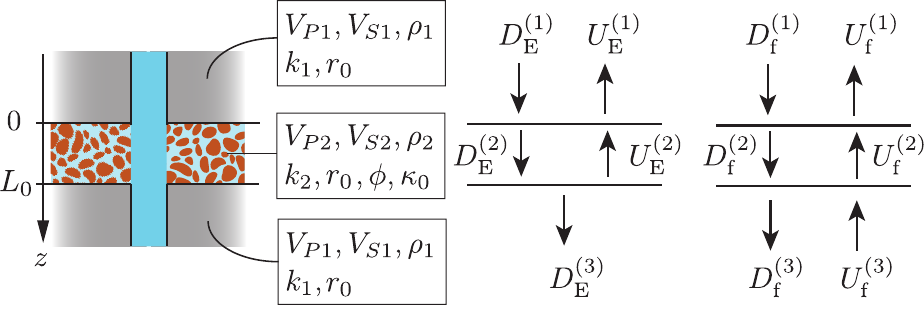}
\caption{A borehole with a constant radius ($r_0$) in a porous formation with thickness $L_0$ sandwiched between two elastic half-spaces with identical properties. $U^{(i)}_{\rm E}$ and $D^{(i)}_{\rm E}$ are upgoing and downgoing elastic-wave potential amplitudes, and $U^{(i)}_{\rm f}$ and $D^{(i)}_{\rm f}$ tube-wave potential amplitudes. }
\label{fig:Geom_TPL}
\end{figure}

In this subsection, we consider tube waves generated at a poroelastic layer sandwiched between elastic layers. The thickness of the poroelastic layer is thin compared to the seismic wavelength (Figure \ref{fig:Geom_TPL}). \citeA{Li1994} earlier studied a similar scenario, where the tube-wave amplitude is derived based on the fluid-continuity equation within a porous layer and the fluid volume at the intersection between the layer and the borehole (see \ref{Append:Eff_Src} for more details). We discuss the relation between the theory developed in this study and other models in terms of consistency with the numerical solutions of the Biot theory. 

We investigate a stack of three layers ($N=3$) with two boundaries in equation \ref{eq:unu1}. We assume that the top layer ($n=1$) and the bottom layer ($n=3$) are elastic layers with identical material properties, and the middle layer ($n=2$) is a porous formation with thickness $L_0$ (Figure \ref{fig:Geom_TPL}). The contribution of fluid infiltration in the source term ($\Delta p_{\rm ft}$ and $\Delta v_{\rm ft}$ in equations \ref{eq:delta_p} and \ref{eq:delta_v}) is non-zero only in the middle layer ($n=2$). The radiation conditions at the upper ($z=z_1$) and lower ($z=z_2$) boundaries are the same as in equations \ref{eq:Df1_2L} and \ref{eq:Uf2_2L}, respectively, except that an additional phase-shift term at the lower boundary is necessary for equation \ref{eq:Uf2_2L} due to non-zero depth ($z_2=L_0$). Similar to the two-layer model (Section \ref{Sec:Elastic}), the borehole pressure waveform can be obtained by solving equation \ref{eq:unu1} and evaluating equation \ref{eq:p}.

\subsubsection{Verification through finite-difference modeling}
We first verify our simplified theory using FD modeling results of Biot dynamic poroelasticity. We consider a poroelastic layer with layer thickness 1 m ($L_0=1$ m), permeability 1 darcy ($\kappa_0=9.869\times 10^{-13}$ $\rm m^2$), and porosity 0.3 ($\phi=0.3$). All material properties are summarized in Table \ref{table:TPL}. The grain bulk modulus ($K_S$), frame bulk modulus ($K_m$), frame shear modulus ($\mu$), and grain density ($\rho_s$) of the poroelastic layer in FD modeling are selected such that their low-frequency elastic limits (\ref{Append:B}) are identical to those of the surrounding elastic half-spaces: the tube waves are generated only due to fluid infiltration from the poroelastic layer. We use the tortuosity factor ($\mathcal{T}$) considered in \citeA{Ou2019GJI}, which assumes round pores. 

\begin{figure}
\centering
 \noindent\includegraphics{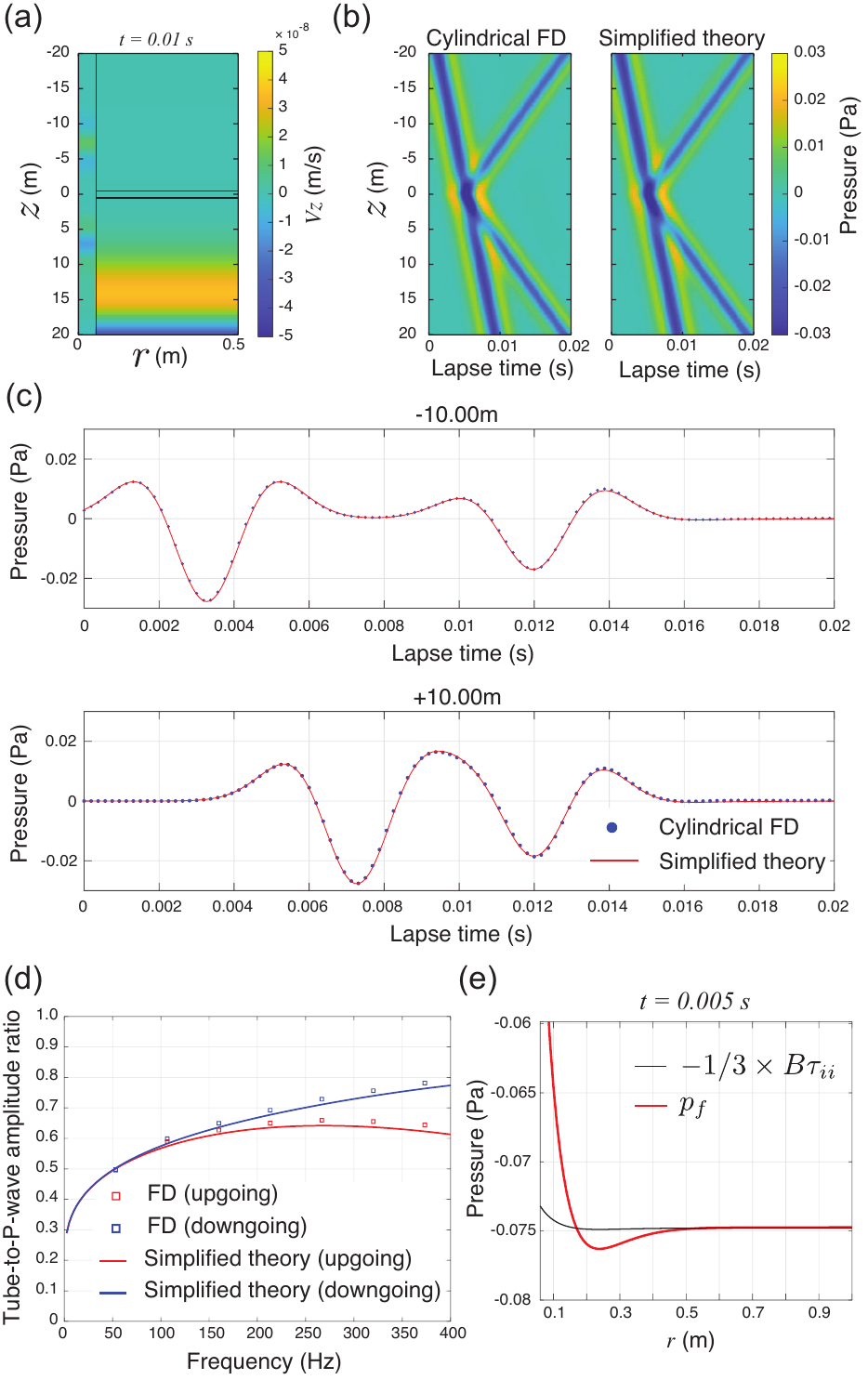}
\caption{Verification using FD modeling for a borehole through a thin porous formation (Figure \ref{fig:Geom_TPL} and Table \ref{table:TPL}) with thickness $L_0$ = 1 m. (a) A snapshot of the wavefield calculated by cylindrical (poroelastic) FD. (b) Comparison of the pressure waveforms in the borehole calculated by FD and the theory developed in this study. (c) Same as (b) but for a receiver located at $\pm 10.00$ m distance from the center of the poroelastic layer. (d) Tube-to-P-wave amplitude ratio at each frequency. (e) Distribution of pore pressure $p_f$ (red line) at the center of the poroelastic layer calculated by FD at the lapse time where the P wave propagates through the layer. The black line shows pore pressure assuming an undrained condition (equation \ref{eq:pf_App}), calculated using the total stress from the FD modeling result.}
\label{fig:TPL}
\end{figure}

Figure \ref{fig:TPL}(a) shows the numerical modeling results (snapshot) using the FD method. The poroelastic layer is located between $z=\pm0.5$m. The incident P wave propagates through the poroelastic layer. It produces tube waves which are visible at $z=\pm7.5$ m in the borehole. In the simplified theory, we use the values of the porous-layer properties from the Gassmann's low-frequency limits (see Table \ref{table:TPL}). A comparison between the FD modeling results and the simplified theory (Figures \ref{fig:TPL}b and \ref{fig:TPL}c) shows an excellent agreement. In Figures \ref{fig:TPL}(b) and \ref{fig:TPL}(c), one can see the tube waves generated due to fluid infiltration from the porous layer. Furthermore, there is no reflected P wave due to the vanishingly small elastic impedance contrast. Contrary to the case of the elastic impedance boundary (Section \ref{Sec:Elastic}), the upgoing and downgoing tube waves have the same polarity. 

Figure \ref{fig:TPL}(d) shows the tube-to-P-wave amplitude ratio. The amplitude ratio from the new, simplified theory (solid lines in Figure \ref{fig:TPL}d) is calculated by identifying the terms that are associated with the generated tube waves in the algebraic expressions of the equation \ref{eq:p} at $z=-Z$ and $z=L+Z$, as demonstrated in the case of an elastic impedance boundary (Section \ref{Sec:Elastic}). The ratio from the FD method (open squares in Figure \ref{fig:TPL}d) is calculated by windowing the incident P wave and the generated tube wave in the time domain, where the tube wave is isolated by subtracting the waveform without the poroelastic layer from the total response. The tube-to-P-wave amplitude ratio shows frequency dependence. The frequency dependence is different between the upgoing and the downgoing tube waves (Figure \ref{fig:TPL}d), which indicates that their time-domain waveforms are not identical. Note that this effect has not been considered in the earlier studies \cite{Li1994}. The difference in the amplitude ratio becomes significant at higher frequencies (above 100 Hz; Figure \ref{fig:TPL}d). This suggests that, for a porous layer, the effect of asymmetric phase interferences in the continuous source located over the layer (equations \ref{eq:delta_pft} and \ref{eq:delta_vft}) grows with growing layer thickness relative to the seismic wavelength. 

Finally, we examine the importance of taking into account the Skempton coefficient in relating the elastic wave pressure to the pore pressure in the newly developed simplified theory (equation \ref{eq:pext}). Figure \ref{fig:TPL}(e) shows pore-pressure distribution at the center of the poroelastic layer ($z=0$ m in Figure \ref{fig:TPL}a) along the radial direction calculated by the FD method at $t=0.005$ s. The black line indicates a value of $-1/3\times B\tau_{ii}$, where $\tau_{ii}$ is the trace of the total stress tensor in the Biot theory (see \ref{Append:Biot}), calculated by the FD method. This illustrates that the pore pressure away from the borehole converges to $-1/3\times B\tau_{ii}$. The Skempton coefficient in equation \ref{eq:pext} is necessary in order to be consistent with the result of the Biot theory. 

 \begin{table}
 \caption{Material properties used in calculating the fluid-pressure response in the borehole at a three-layer model including a thin porous layer (Figure \ref{fig:Geom_TPL}). The properties used in the finite-difference (FD) modeling of Biot dynamic poroelasticity and those in the simplified theory developed in this study are shown.}
 \centering
 \begin{tabular}{cll}
     & FD & Simplified theory (this study) \\
 \hline
\begin{tabular}{c}
    Porous layer \\
    (middle)
\end{tabular} & 
\begin{tabular}{l}
$\kappa_0$=1 darcy, $\phi$=0.3, \\
$K_S$=100 GPa, $K_m$=28.8 GPa,  \\ 
$\mu$=22.5 GPa, $\rho_s$=3.14 g/cc, \\
$\mathcal{T}$=3.75
\end{tabular}  & 

\begin{tabular}{l}
$\kappa_0$=1 darcy, $\phi$=0.3, \\ $V_{P}$=5 km/s,
$V_{S}$=3 km/s,\\ $\rho$=2.5 g/cc 
\end{tabular} \\
\hline
\begin{tabular}{c}
    Elastic layers \\
    (top and bottom)
  \end{tabular} &   \multicolumn{2}{c}{$V_{P}$=5 km/s, $V_{S}$=3 km/s, $\rho$=2.5 g/cc} \\
\hline
 \end{tabular}
\label{table:TPL}
 \end{table}

\subsubsection{Thin-layer approximation and relation to other models}
Next, we show the closed-form expression of the amplitude using the simplified theory and discuss its relevance to the previous studies \cite{Li1994}. To consider the situation where the porous layer is thin compared to the seismic wavelength \cite{Li1994}, we derive a first-order approximation of the analytical amplitudes in terms of layer thickness $L_0$. This is achieved by obtaining the pressure response at $z=-Z$, and $z=L_0+Z$, respectively, and evaluating their Taylor series expansion in terms of $L_0$. After tedious algebra, we obtain the following equation for the approximated borehole pressure response at the top layer ($z=-Z$):
\begin{align}
     p(-Z) \approx& -\rho_f C_T^{(1)} \omega\frac{2 k_1 k_{p_1}}{k_{p_1}^2-k_1^2}A^{(1)}_P D_{\rm E}^{(1)} e^{i k_{p_1}(-Z)} \nonumber \\
            & -\rho_f C_T^{(1)} \omega\frac{2 k_1 k_{p_1}}{k_{p_1}^2-k_1^2}A^{(1)}_P \tilde{R} D_{\rm E}^{(1)} e^{-i k_{p_1}(-Z)}  \nonumber \\
            & + \tilde{A}_{\uparrow} e^{-i k_{1}(-Z)}, \label{eq:p_tpl_up}
\end{align}
where each term on the right-hand side represents the direct P wave, reflected P wave, and generated tube wave. The approximate P-wave reflection coefficient $\tilde{R}$ is given by
\begin{equation}
     \tilde{R} = -\frac{i}{2}\frac{k_{p_1}^2\rho_2^2-k_{p_2}^2\rho_1^2}{\rho_1\rho_2 k_{p_1}}L_0. \label{eq:R_aprx}
\end{equation}
The approximate upgoing tube wave amplitude $\tilde{A}_{\uparrow}$ is composed of the following three terms:
\begin{equation}
     \tilde{A}_{\uparrow} = \tilde{A}_{\uparrow}^{E}+\tilde{A}^{\Delta k}+\tilde{A}^{B}, \label{eq:Aup_aprx}
\end{equation}
where the first term ($\tilde{A}_{\uparrow}^{E}$) contains the contribution due to the difference in the elastic properties ($A_P$, $\rho$, and $k_p$), the second term ($\tilde{A}^{\Delta k}$) the contribution due to the difference in the tube-wave wavenumber ($k$), and the third term ($\tilde{A}^{B}$) the contribution of fluid infiltration associated with the Skempton coefficient at the porous formation:
\begin{align}
& \tilde{A}_{\uparrow}^{E} = -i\omega\rho_f C_T^{(1)} k_{p_1} \frac{\left(-A_P^{(1)} k_1 \rho_1 \rho_2 + A_P^{(1)} k_1 \rho_2^2 + A_P^{(2)} k_{p_2} \rho_1^2\right)k_{p_1} - A_P^{(1)}k_{p_2}^2\rho_1^2}{\rho_1\rho_2 (k_{p_1}^2-k_{1}^2)}D_{\rm E}^{(1)}L_0, \label{eq:Aup_aprx_E} \\
& \tilde{A}^{\Delta k} = -i\omega\rho_f C_T^{(1)} \frac{A_P^{(1)}k_{p_1}\rho_2k_2^2 -A_P^{(2)}k_{p_2}\rho_1k_1^2 }{\rho_2 (k_{p_1}^2-k_{1}^2)}D_{\rm E}^{(1)}L_0, \label{eq:Aup_aprx_Dk} \\
& \tilde{A}^{B} = -i\omega\rho_f C_T^{(1)} \frac{k_{p_2}^2 \rho_1 \phi \Phi K B}{\rho_2 K_f}D_{\rm E}^{(1)}L_0. \label{eq:Aup_aprx_B} 
\end{align}
Note that the difference in the tube-wave wavenumber ($k$) is caused by the fluid infiltration (inclusion of the porous layer) as well as the difference of the shear modulus between the porous layer and the surrounding elastic layers (see equation \ref{eq:Keff}).  

Similarly, the approximated response at the bottom layer ($z=L_0+Z$) becomes,
\begin{align}
     p(L_0+Z) \approx& -\rho_f C_T^{(1)} \omega\frac{2 k_1 k_{p_1}}{k_{p_1}^2-k_1^2}A^{(1)}_P \tilde{T} D_{\rm E}^{(1)} e^{i k_{p_1}(L_0+Z)} \nonumber \\
            & + \tilde{A}_{\downarrow} e^{i k_{1}(L_0+Z)}, \label{eq:p_tpl_low}
\end{align}
where each term indicates the transmitted P wave and generated tube wave. The approximate transmission coefficient $\tilde{T}$ and the approximate downgoing tube wave amplitude $\tilde{A}_{\downarrow}$ are written as,
\begin{align}
     \tilde{T} &= 1+\frac{i}{2}\frac{-k_{p_1}^2\rho_1\rho_2+k_{p_1}^2\rho_2^2+k_{p_2}^2\rho_1^2}{\rho_1\rho_2 k_{p_1}}L_0 \label{eq:T_aprx}, \\
     \tilde{A}_{\downarrow} &= \tilde{A}_{\downarrow}^{E}+\tilde{A}^{\Delta k}+\tilde{A}^{B}. \label{eq:Adown_aprx}
\end{align}
The downgoing amplitude $\tilde{A}_{\downarrow}$ is the same as the upgoing amplitude (equation \ref{eq:Aup_aprx}) except for the contribution due to the difference in the elastic properties $\tilde{A}_{\downarrow}^{E}$: 
\begin{equation}
 \tilde{A}_{\downarrow}^{E} = -i\omega\rho_f C_T^{(1)} k_{p_1} \frac{\left(A_P^{(1)} k_1 \rho_1 \rho_2 - A_P^{(1)} k_1 \rho_2^2 + A_P^{(2)} k_{p_2} \rho_1^2\right)k_{p_1} - A_P^{(1)}k_{p_2}^2\rho_1^2}{\rho_1\rho_2 (k_{p_1}^2-k_{1}^2)}D_{\rm E}^{(1)}L_0. \label{eq:Adown_aprx_E}
\end{equation}
It is clear from equations \ref{eq:Aup_aprx_E} and \ref{eq:Adown_aprx_E} that the generated tube waves become identical (i.e., $\tilde{A}_{\uparrow}=\tilde{A}_{\downarrow}$) when $\rho_2=\rho_1$. 

\begin{figure}
\centering
 \noindent\includegraphics{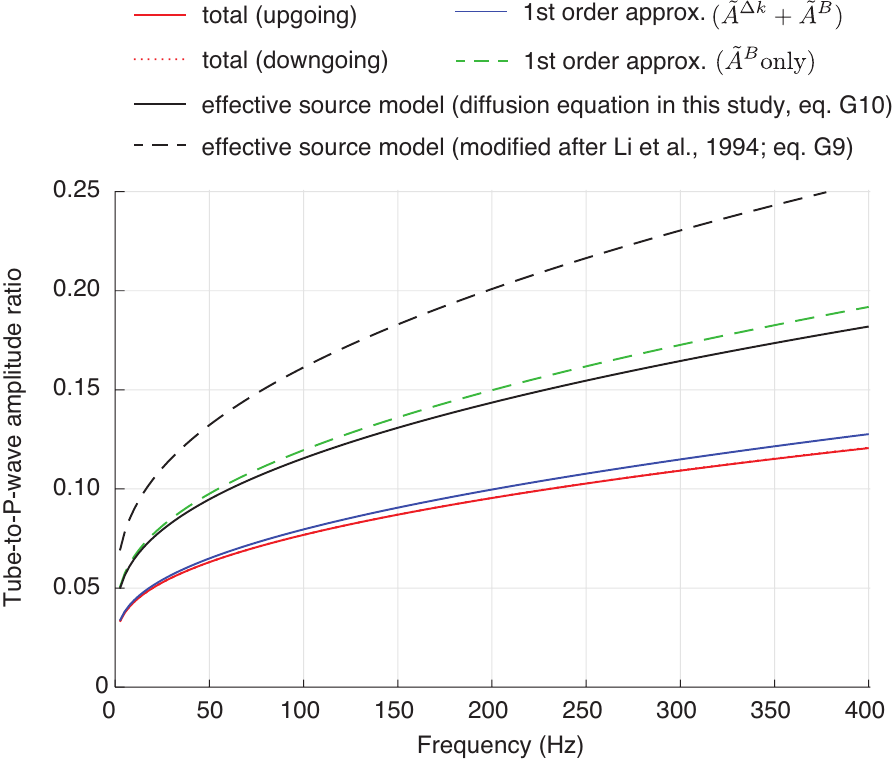}
\caption{Comparison of tube-to-P-wave amplitude ratio calculated using various tube-wave generation models for a thin porous layer sandwiched between two elastic layers (Figure \ref{fig:Geom_TPL}) when the layer thickness $L_0$ = 0.1 m. Note that in first-order approximation, the term associated with the elastic impedance contrast is zero (i.e., $\tilde{A}_{\uparrow}^{E}$ = $\tilde{A}_{\downarrow}^{E}$ = 0 in equations \ref{eq:Aup_aprx} and \ref{eq:Adown_aprx}) because of the material properties considered (Table \ref{table:TPL}). ``total'': using the theory developed in this study and solved by the propagator matrix method (equations \ref{eq:unu1} and \ref{eq:p}). Solid red (upgoing) and dashed red (downgoing) lines are on top of each other. ``1st order approx'': the first-order approximation of the total response in terms of $L_0$ (equations \ref{eq:p_tpl_up} to \ref{eq:Adown_aprx_E}). ``effective source model'': the amplitude based on the flow volume and the diffusion equation (\ref{Append:Eff_Src}).}
\label{fig:Eff_Src}
\end{figure}

Next, the analytical amplitudes from different formulations are compared (Figure \ref{fig:Eff_Src}). In the past, the analytical amplitudes have been derived based on the effective-source formulation for porous layers \cite{Li1994,Minato2017SEG} and open fractures \cite{Ionov2007,Bakku2013,Minato2017JGR}. The effective-source model uses a fluid-continuity equation at a permeable structure (porous layer or open fracture) and relates the fluid volume with the tube-wave amplitude (\ref{Append:Eff_Src}). Figure \ref{fig:Eff_Src} shows the tube-to-P-wave amplitude ratio for the three-layer model with properties presented in Table \ref{table:TPL} and assuming a small layer thickness, $L_0=0.1$ m. Also shown is the amplitude ratio using the earlier model of \citeA{Li1994}, modified using the effective-source formulation (\ref{Append:Eff_Src}). For the sake of completeness, furthermore, we also derive the effective-source model using the pore-pressure diffusion equation in the simplified theory (see \ref{Append:Eff_Src}). 

The upgoing and downgoing tube waves of the total solutions without the first-order approximation (solid red and dashed red lines in Figure \ref{fig:Eff_Src}) are almost identical, and they are on top of each other. The effective-source models (solid black and dashed black lines in Figure \ref{fig:Eff_Src}) show a large deviation from the total solution even at very low frequencies. In contrast, the first-order approximation (blue line; equation \ref{eq:Aup_aprx} or equation \ref{eq:Adown_aprx}) reasonably represents the total solution. The green dashed line in Figure \ref{fig:Eff_Src} shows the contribution due to the fluid infiltration ($\tilde{A}^{B}$) ignoring the difference in the tube-wave wavenumber ($\tilde{A}^{\Delta k}$). In this case, the first-order approximation converges at low frequencies to the effective-source model (see dashed green and solid black lines in Figure \ref{fig:Eff_Src}). This suggests that the effective-source formulation does not take into account the tube-wave velocity ($C_T$) at the porous layer that is locally different from $C_T$ at the surrounding elastic formation (see equation \ref{eq:Keff}). In this vein, it is well known that the propagating tube waves experience significant reflection and transmission effects at a boundary where tube-wave velocity markedly changes \cite<e.g.,>{Tezuka1997}. Therefore, when formulating the tube-wave amplitude based on the effective-source model, it would be necessary to consider if the generated tube wave is scattered immediately after generation due to a difference in the tube-wave velocity with respect to the surrounding formation. Similar discussion, but in the context of the boundary condition for open fractures, can be found in \citeA{Minato2017JAP}. 

The effective-source model based on the earlier studies of \citeA{Li1994} predicts a larger amplitude ratio than the effective-source model using the diffusion equation presented in this study (see solid black and dashed black lines in Figure \ref{fig:Eff_Src}). This difference is caused by their diffusion equation incompatible with ours (equation \ref{eq:Li_PDE}) and the drained bulk modulus approximated by the Reuss average (see \ref{Append:Eff_Src}).

\subsection{Step-like change in borehole radius}
\label{Sec:Caliper}

The last example is of borehole irregularities modeled as changes in the borehole radius (Figure \ref{fig:intro_3M}c). Although the closed-form expression of the generated tube-wave amplitude is known \cite{White1988}, we derive it from our newly developed theory; we also show a comparison with the FD modeling results. 

\begin{figure}
\centering
 \noindent\includegraphics{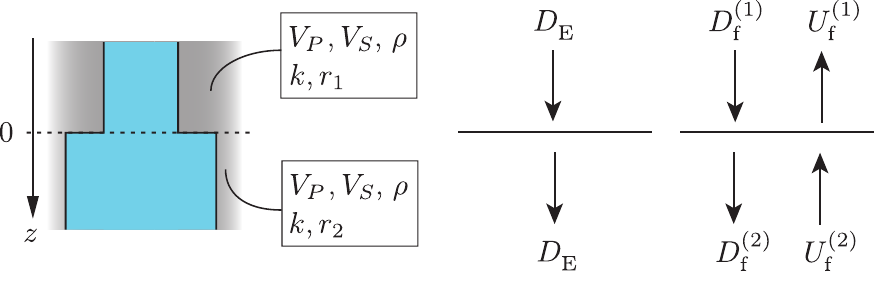}
\caption{A borehole with an irregular radius located in a homogeneous elastic formation. There is only the incident elastic wave $D_{\rm E}$; $U^{(i)}_{\rm f}$ and $D^{(i)}_{\rm f}$ are upgoing and downgoing tube-wave potential amplitudes. }
\label{fig:Geom_Caliper}
\end{figure}

We test on a two-layer model with a single boundary located where the borehole radius changes stepwise (Figure \ref{fig:Geom_Caliper}). The surrounding elastic medium is homogeneous. In this case, equation \ref{eq:unu1} becomes
\begin{linenomath*}
 \begin{equation}
    \begin{pmatrix}
    U^{(1)}_{\rm f} \\
    D^{(1)}_{\rm f}
   \end{pmatrix}=\mathbf{M}_1(0)
    \begin{pmatrix}
    U^{(2)}_{\rm f} \\
    D^{(2)}_{\rm f}
   \end{pmatrix}+\frac{\Delta v_{\rm q}}{2\omega k}
    \begin{pmatrix}
    1 \\
    -1
   \end{pmatrix}.
  \label{eq:un_2L_caliper}
 \end{equation}
\end{linenomath*}
The pressure response can be calculated in the same manner as in Section \ref{Sec:Elastic}:
\begin{align}
     p(\pm Z) =& -\rho_f C_T \omega\frac{2 k k_{p}}{k_{p}^2-k^2}A_P D_{\rm E} e^{i k_{p}(\pm Z)} \nonumber \\
            & + A_{C} e^{\pm i k (\pm Z)}, \label{eq:p_caliper_low}
\end{align}
where the first term is the direct P wave, and the second term is the generated tube wave. The amplitude of the tube wave can be written as
\begin{equation}
     A_{C}=-\rho_f C_T \omega\frac{k_{p}}{k_{p}^2-k^2}\left(\frac{r_1^2-r_2^2}{r_1^2+r_2^2}\right)\left(2 A_P k_p-k^2+k_p^2\right)D_{\rm E}. \label{eq:A_caliper}
\end{equation}
Equation \ref{eq:A_caliper} is identical to the known solution using the quasi-static approximation \cite<equation 4 in>[]{White1988}.

\begin{figure}
\centering
 \noindent\includegraphics{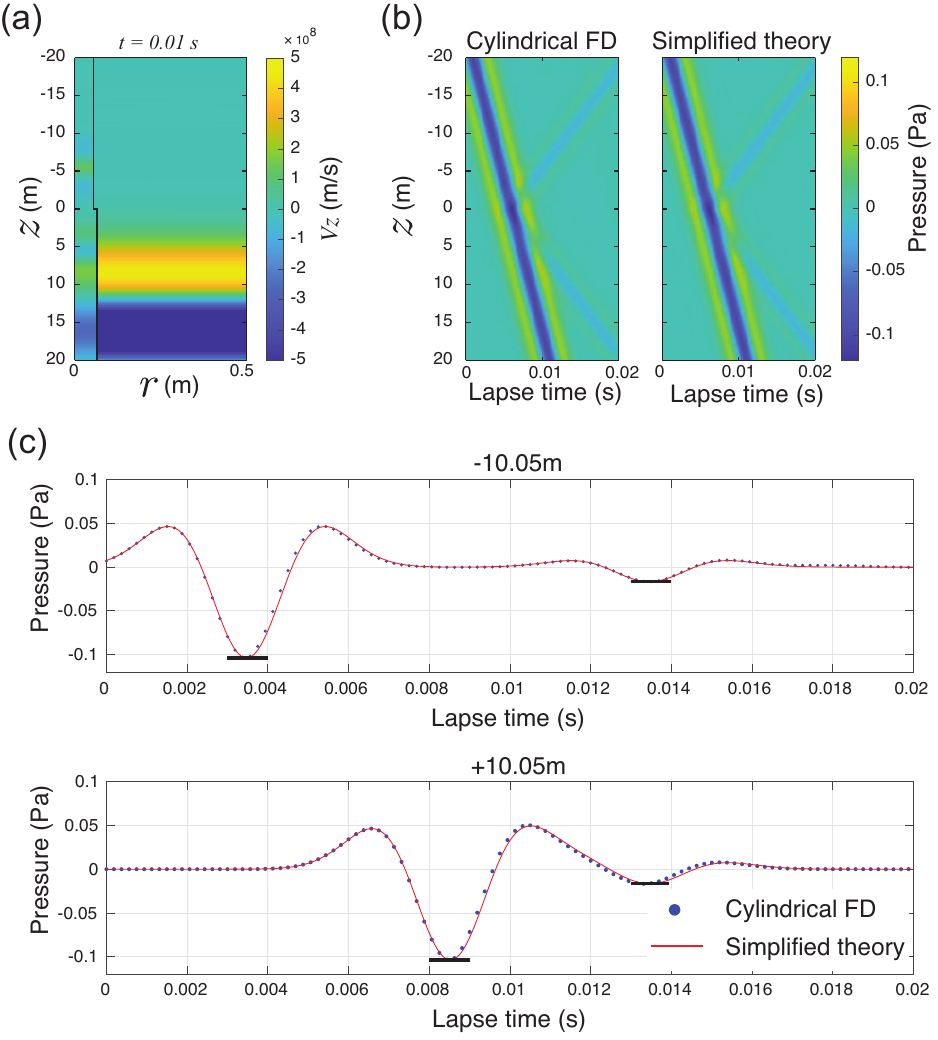}
\caption{Verification using FD modeling for an irregular borehole embedded in a homogeneous elastic medium (Figure \ref{fig:Geom_Caliper}). (a) A snapshot of wavefield calculated by cylindrical FD. (b) Comparison of pressure waveforms at the borehole calculated by FD and those calculated by the theory developed in this study. (c) Same as (b) but for a receiver located at $\pm 10.05$ m from the boundary where the radius changes. The red lines are calculated waveforms using our theory and the propagator matrix method (equations \ref{eq:unu1} and \ref{eq:p}). The black lines are the analytical amplitudes derived from the same theory (equations \ref{eq:p_caliper_low} and \ref{eq:A_caliper}). }
\label{fig:Caliper}
\end{figure}

Figure \ref{fig:Caliper} presents a comparison between the simplified theory and the FD modeling results. We consider that the radius at the upper half-space ($r_1$) is 0.055 m, and that at the lower half-space ($r_2$) is 0.065 m. These results show excellent agreement between the simplified theory and the FD modeling results. 

\section{Hydrophone VSP experiment at a fault damage zone}

\subsection{General experimental settings}
\label{Sec:Field_Geology}

\begin{figure}
\centering
 \noindent\includegraphics[width=\textwidth]{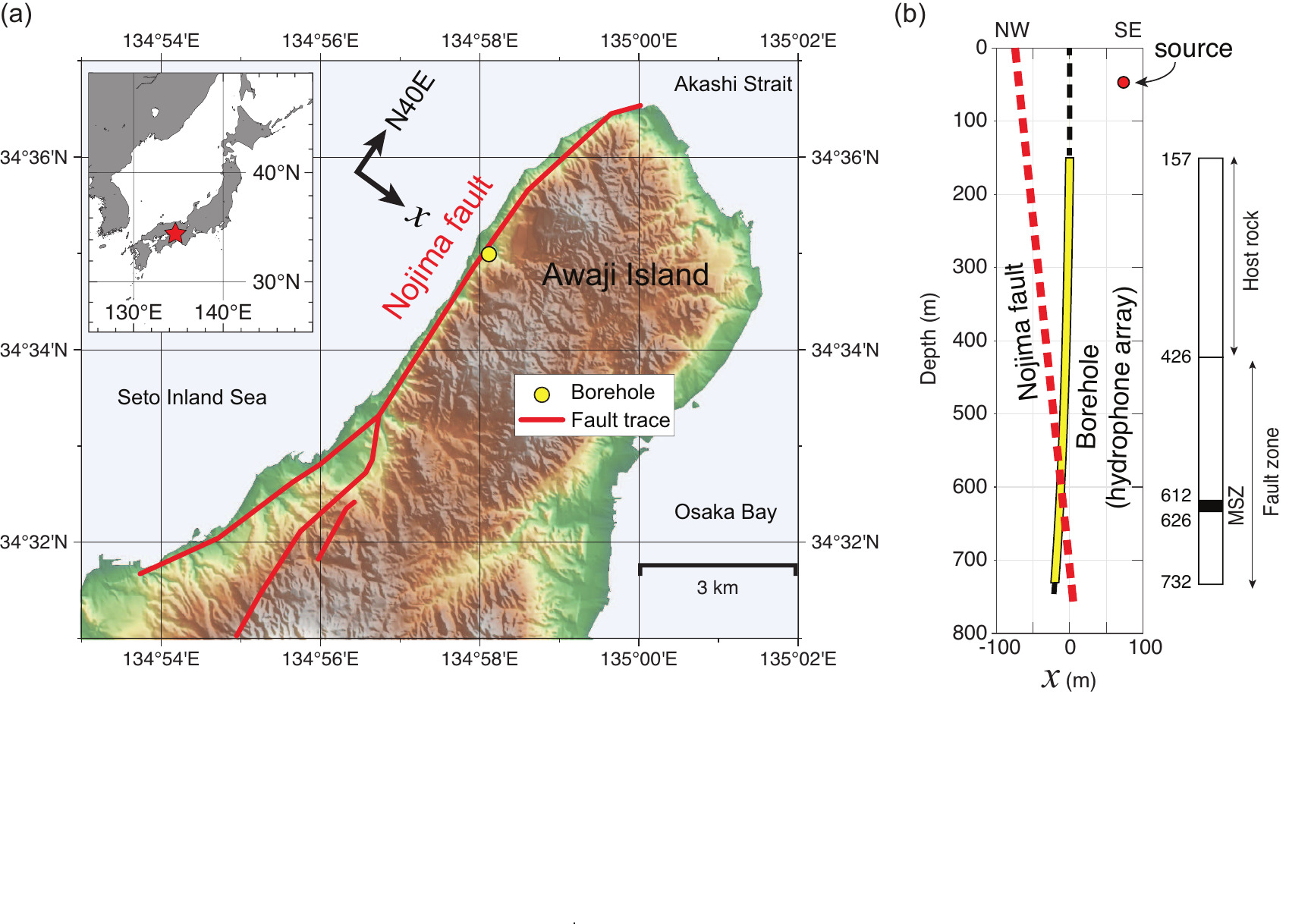}
\caption{(a) Location of the measurement borehole and the track of the Nojima fault in Awaji Island, southwest Japan. (b) Trajectory of the fault, the depth of the hydrophone measurements used in this study, and the location of the seismic source for the VSP experiment. All of them are projected in a plane perpendicular to the fault strike. The depth of the host rock region, the fault damage zone, and the main shear zone (MSZ) were derived from core analyses \cite{Tanaka2001}.}
\label{fig:Awaji}
\end{figure}

A hydrophone VSP experiment was performed at the Nojima fault in Awaji Island, southwest Japan (Figure \ref{fig:Awaji}a), as a part of a project aimed at scientific drilling in active fault localities \cite{Ito1999}. A borehole was drilled from the surface to 746.7 m depth. The borehole is located at approximately 74.6 m distance from the surface exposure of the Nojima fault which ruptured during the Mw 6.9 1995 Kobe earthquake. Extensive core analyses \cite<e.g.,>[]{Fujimoto2001,Ohtani2001,Tanaka2001} and in-situ geophysical measurements \cite<e.g.,>[]{Ito1996,Roeloffs1999,Kiguchi2001,Ito2005} have been performed in this borehole. From the surface fault rupture and the borehole observations, the dip of the Nojima fault is inferred to be approximately 83$^\circ$; the fault surface intersects the borehole approximately at 625 m depth \cite{Tanaka2001}. The hydrophone VSP experiments were performed using dynamite sources placed at 47 m depth and at a lateral distance of 96.1 m from the borehole. We use data measured between 157 m and 732 m depth with hydrophone spacing of 1 m in an uncased, open section of the borehole (Figure \ref{fig:Awaji}b). More details about the data acquisition can be found in \citeA{Kiguchi2001}. From borehole observation and core analyses \cite{Tanaka2001}, the survey depth until 426 m is identified as the host rock region including granodiorite and porphyry intrusive bodies. The depth below 426 m is identified as the fault damage zone consisting of weakly pulverized and altered rocks (WPAR), cataclasite, and fault breccia. The depth 612--626 m is the main shear zone (MSZ) consisting of ultracataclasite and pseudotachylite (Figure \ref{fig:Awaji}b). 

The measured hydrophone data and the downhole logging data at the same depth interval are shown in Figure \ref{fig:intro_data}. A vertical hydrophone array of 18 m length was sequentially moved in the borehole in order to cover the survey depth. The measured amplitudes were corrected for the variations in source energy as monitored at the surface. Figure \ref{fig:intro_data}(a) displays the data after a depth-dependent scaling for better visibility. The downhole log data (Figure \ref{fig:intro_data}b) consist of the sonic velocities ($V_P$, $V_S$) from dipole shear sonic imager (DSI), the gamma-gamma density, the radius from the caliper log, and the neutron porosity.

\subsection{Data preprocessing}
\label{Sec:Field_Preproc}
\begin{figure}
\centering
 \noindent\includegraphics{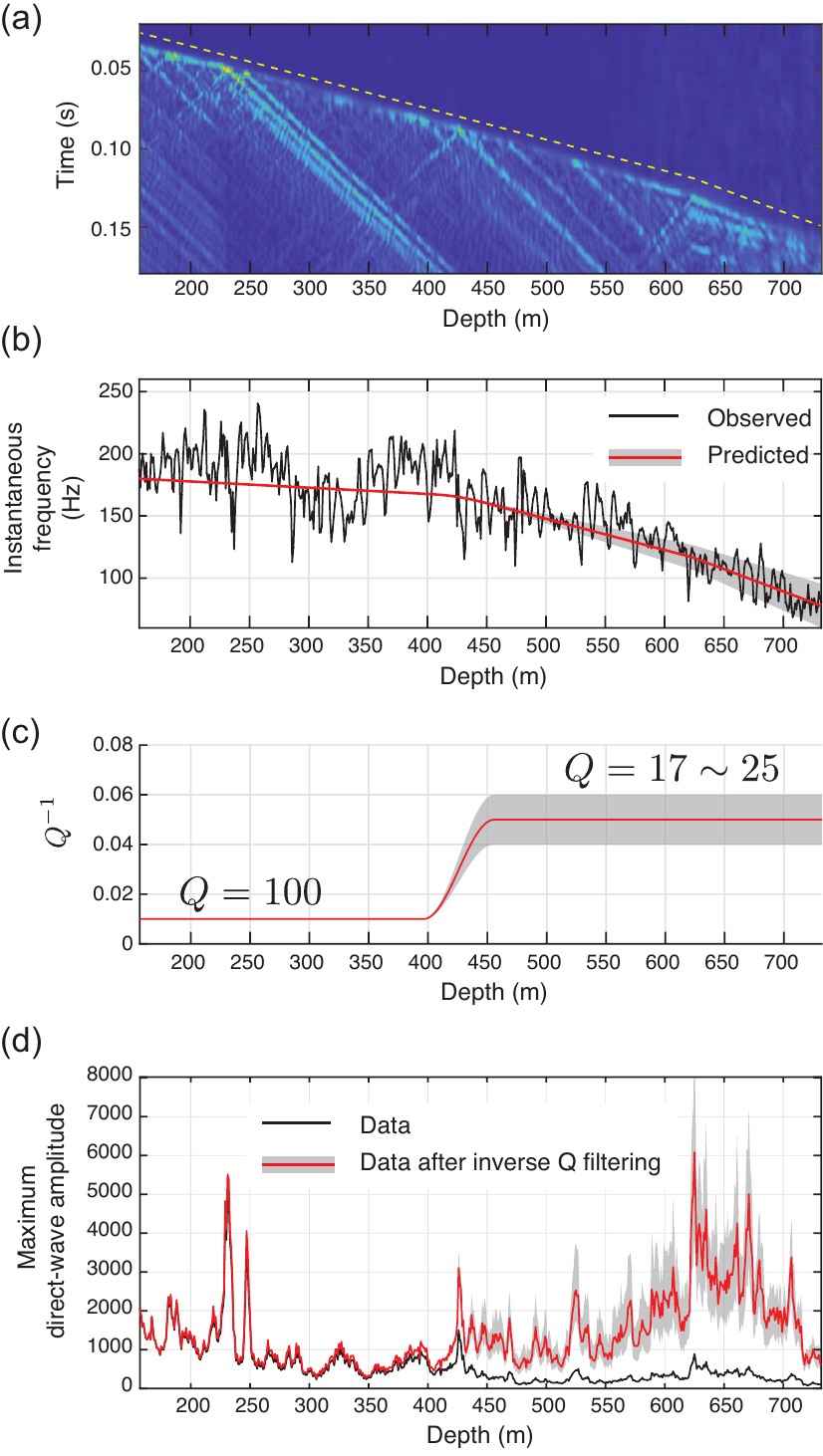}
\caption{Result of attenuation ($Q$ factor) analyses. (a) Envelope of the observed hydrophone data (Figure \ref{fig:intro_data}a). The yellow dashed line indicates the traveltime assuming a two-layer model. This traveltime is used in the geometrical-spreading correction. (b) Instantaneous frequency at the envelope maxima of the direct-wave waveform in (a). The red line indicates the calculated centroid-frequency shift using the distribution of $Q$ shown in (c). (c) Assumed $Q$ values as a function of depth. (d) Maximum pressure amplitude of the direct-wave waveforms after geometrical-spreading correction (black line) and the same after additional attenuation correction (red line) using $Q$ values shown in (c).}
\label{fig:Q}
\end{figure}

In order to discuss the amplitude distribution of the field pressure response, we first analyze the P-wave amplitude attenuation or the quality ($Q$) factor in the observed data. We estimate the $Q$ factor using the approach of \citeA{Vesnaver2020}. This approach assumes that the instantaneous frequency at the maxima of the envelope of a direct wave represents the spectral centroid of the signal. The shifts in the centroid frequency are then utilized to measure $Q$ \cite{Quan1997}. The attenuation analyses of \citeA{Vesnaver2020} are especially advantageous to our dataset because the conventional methods that involve time-windowing of direct P wave \cite<e.g., the spectral ratio method,>[]{Tonn1991} may be significantly affected by the interference of tube waves. Figures \ref{fig:Q}(a) and \ref{fig:Q}(b) show the envelope of the measured hydrophone data and the calculated instantaneous frequencies. The observed instantaneous frequencies range from 90 Hz to 200 Hz with relatively large fluctuations; the average value decreases with depth. The red line in Figure \ref{fig:Q}(b) indicates the predicted centroid frequency, using the distribution of $Q$ as shown in Figure \ref{fig:Q}(c). Presumably, the P-wave attenuation increases ($Q$ becomes smaller) in the fault damage zone ($Q$ = 17--25 at depths larger than 426 m) than in the host rock region ($Q$ = 100 at depths less than 426 m). The predicted frequency shift reveals the long-wavelength trend in the observed data reasonably well (Figure \ref{fig:Q}b). The details of the attenuation analyses can be found in \ref{Append:Source_Q}.

The distribution of the maximum amplitude of the direct waves in the hydrophone data is shown in Figure \ref{fig:Q}(d). The solid black line in Figure \ref{fig:Q}(d) is the amplitude after compensating for the geometrical spreading; this permits the assumption of a plane-wave incidence. The geometrical-spreading correction is based on the approach of \citeA{Harris1997} which uses the first arrival traveltime (yellow dashed line in Figure \ref{fig:Q}a). The red line in Figure \ref{fig:Q}(d) illustrates the amplitude after compensating for additional attenuation with $Q$ values shown in Figure \ref{fig:Q}(c). To do this, we apply the inverse Q filter \cite{Wang2014} to time-windowed direct waves. Due to significant attenuation, the result after inverse Q filtering shows large amplitudes at depths exceeding 426 m.

\subsection{Role of the three different tube-wave generation mechanisms}
\label{Sec:Field_Synth}

We use the theory developed in this study to calculate the borehole pressure response at the Nojima fault zone. We create a multi-layer model with a constant layer thickness of 0.2 m to represent the medium from 157 m to 732 m depth. The material properties at each layer are assigned by interpolating the downhole log data (Figure \ref{fig:intro_data}b). In calculating the pressure response, the source wavelet and the normal stress associated with the incident P wave are estimated from the observed data (see \ref{Append:Source_Q}). As in the previous section (Section \ref{Sec:Simple_Config}), we consider the radiation conditions of an infinite elastic half-space at the topmost and bottommost layers. 

\begin{figure}
\centering
 \noindent\includegraphics[width=\textwidth]{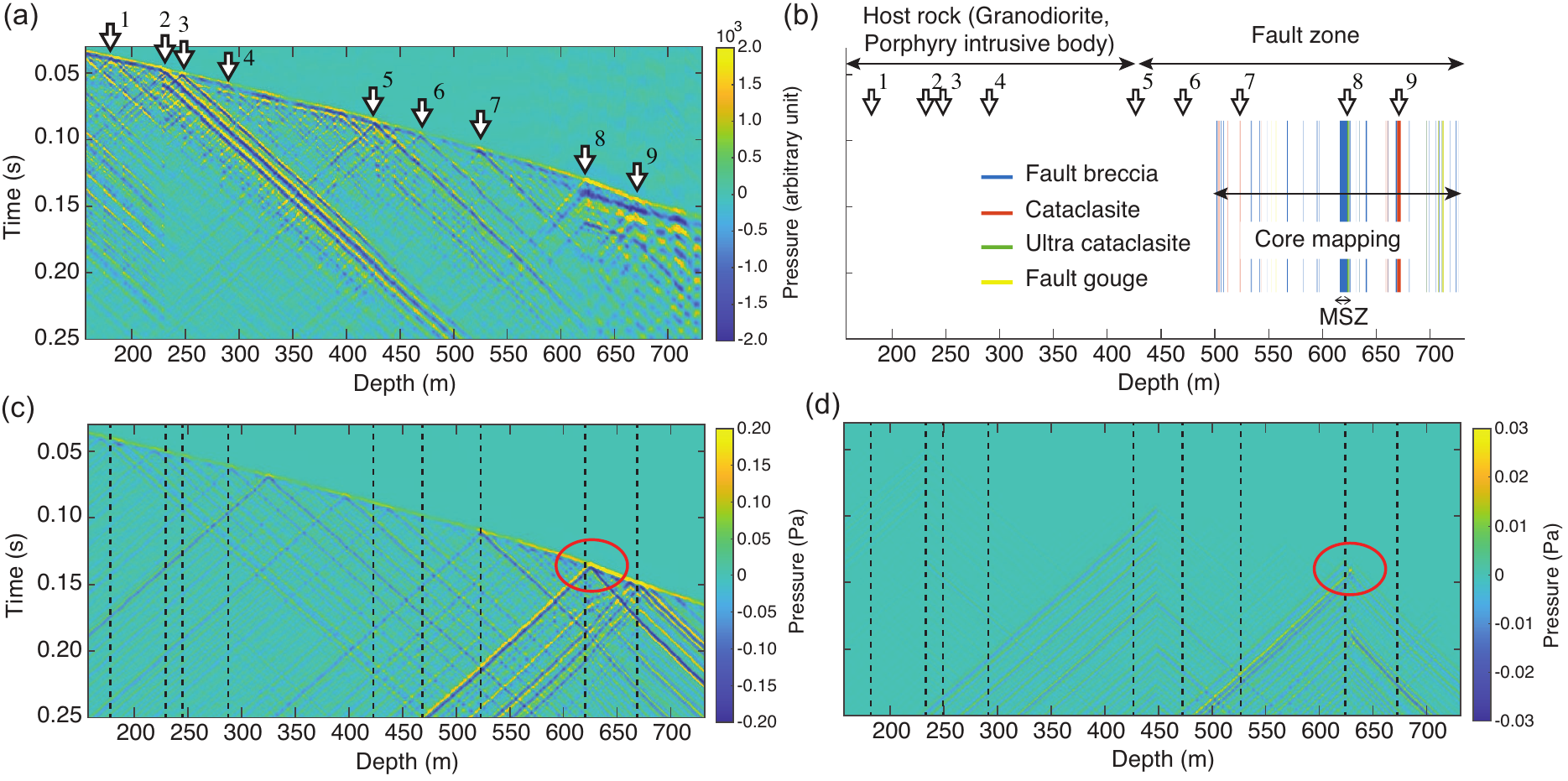}
\caption{(a) Observed pressure response (hydrophone data). Arrows mark the location of the nine permeable structures (Table \ref{table:PZ}). (b) Summary of the depth of fault-related rocks from core analyses \cite{Tanaka2001}. (c) Calculated pressure response using the theory developed in this study and downhole logging data (Figure \ref{fig:intro_data}b), assuming absence of any porous formation. The red ellipse indicates tube waves generated at the depth of the main shear zone (MSZ). (d) Difference waveforms with and without the borehole irregularities, thus including the contribution of the borehole radius changes as seen in the caliper log.}
\label{fig:image_elastic}
\end{figure}

In this study, we focus on tube waves generated at nine depth locations (the arrows in Figure \ref{fig:image_elastic}a). In earlier studies involving tube-wave analyses using the same dataset \cite{Kiguchi2001, Kiguchi2003SEGJ}, the tube waves generated at these depths were interpreted as to be due to open fractures based on the model of \citeA{Beydoun1985}, or porous formations based on the model of \citeA{Li1994}. \citeA{Kiguchi2003SEGJ} estimated the permeability using the tube-to-P-wave amplitude ratio. The location and geological features of those permeable structures are summarized in Table \ref{table:PZ}. The detailed core mapping \cite{Tanaka2001} is available for several permeable structures (i.e., no. 7, 8, and 9 in Table \ref{table:PZ}).  Therefore, Table \ref{table:PZ} also lists the geological features obtained from core analyses. The permeable structures at the nine locations include open fractures in the host rock (granodiorite) and porous formations (e.g., fault breccia and cataclasite) at the main shear zone (Figure \ref{fig:image_elastic}b).

 \begin{table}
 \caption{Permeable structures analyzed in earlier studies at this location \cite{Kiguchi2003SEGJ}.}
 \centering
 \begin{tabular}{cccc}
 \hline
   No.  & Depth (m) &   Geological feature$^a$  & Type of tube-wave model \\
 \hline
   1  & 181--182 & Granodiorite & PZ$^b$ \\
   2  & 232 & Granodiorite & OF$^c$ \\
   3  & 247 & Granodiorite & OF\\
   4  & 290--291 & Granodiorite & PZ\\
   5  & 427 &  WPAR$^d$ & OF \\
   6  & 471 & WPAR & OF \\
   7  & 527 & Cataclasite  & OF\\
   8  & 624--625  & Ultracataclasite & PZ\\
   9  & 669--673  & Fault breccia, Cataclasite & PZ\\
 \hline
\multicolumn{4}{p{\textwidth}}{{\it Note}.  $^a$Geological features of permeable structures: no. 1--6 are from \citeA{Kiguchi2003SEGJ}, no. 7--9 are from the detailed core mapping \cite{Tanaka2001}. $^b$PZ: permeable-zone model \cite{Li1994}. $^c$OF: open-fracture model \cite{Beydoun1985,Hardin1987}. $^d$Weakly pulverized and altered rocks.}
 \end{tabular}
\label{table:PZ}
 \end{table}

We first calculate the pressure response assuming no porous formations (Figure \ref{fig:image_elastic}c); we assume zero permeability ($\kappa_0=0$) at all layers. We also create the same model but having a constant borehole radius. We quantitatively determine the contribution due to the irregular borehole radius by calculating the difference between the response with and without the borehole irregularities (Figure \ref{fig:image_elastic}d). From the magnitude of the calculated amplitudes, we find that the contribution due to borehole irregularities is $\sim$10 \% of the total response; the dominant tube waves in the total pressure response (Figure \ref{fig:image_elastic}c) are generated due to elastic impedance boundaries. Tube wave with the largest amplitude is generated around 624 m depth in the main shear zone or MSZ (red circle in Figure \ref{fig:image_elastic}c). In this zone, the complex structures including fault breccias, ultracataclasites, and pseudotachylites show a significant change in $V_P$, $V_S$, and $\rho$ within a short distance (see Figure \ref{fig:intro_data}b and Figure \ref{fig:image_elastic}b). As demonstrated in Section \ref{Sec:Elastic} and Figure \ref{fig:Elastic_Ratio}, the upgoing and downgoing tube waves generated around MSZ show the polarity difference (red circle in Figure \ref{fig:image_elastic}c). Note that the tube waves generated due to borehole irregularities show the largest amplitude around MSZ (red circle in Figure \ref{fig:image_elastic}d), where the borehole radius increases over a relatively large thickness range ($\sim$15 m) around this depth (Figure \ref{fig:intro_data}b). 

\begin{figure}
\centering
 \noindent\includegraphics{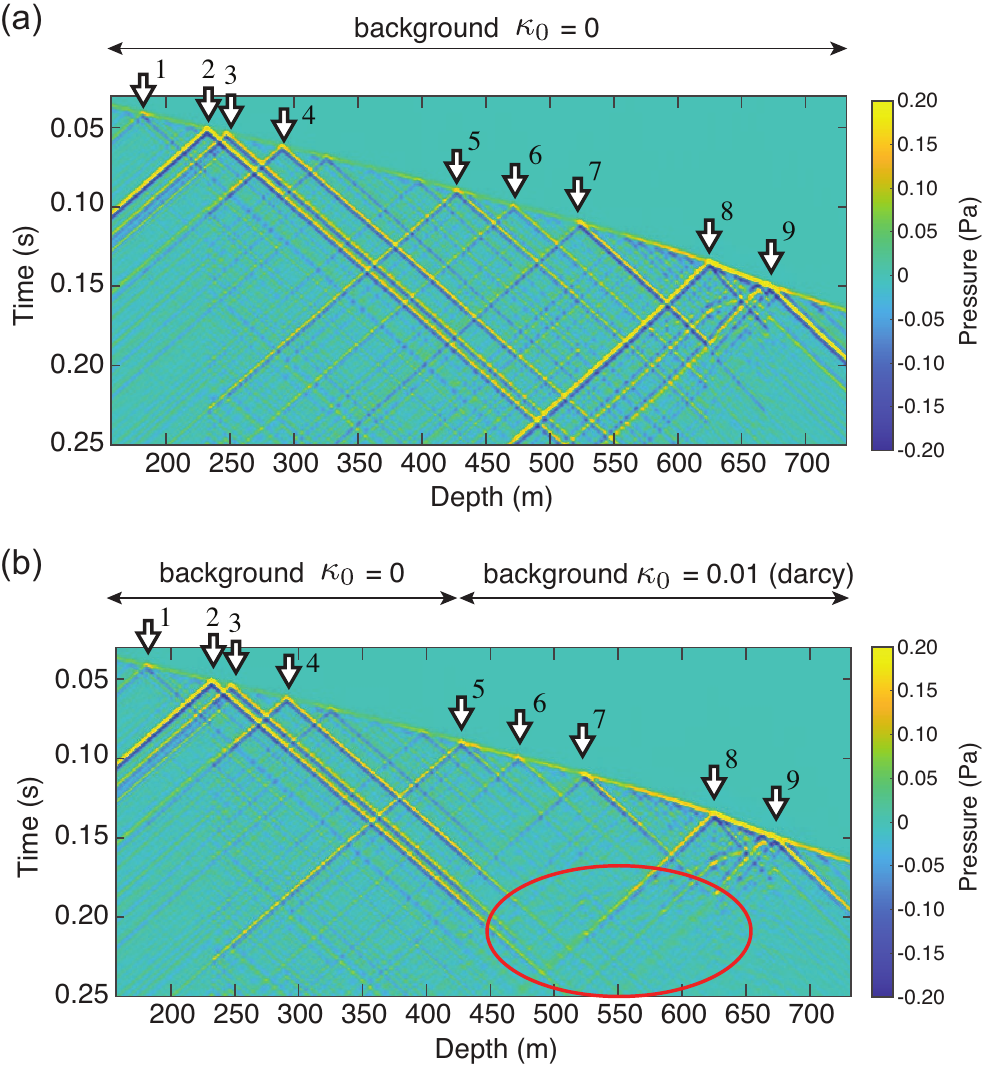}
\caption{(a) Calculated pressure response using the theory developed in this study and downhole logging data (Figure \ref{fig:intro_data}b), assuming the presence of a porous formation at the nine depth locations (Table \ref{table:PZ_synth}). (b) Same as (a) but considering additional non-zero background permeability below 426 m depth (fault damage zone). The red ellipse indicates the area where tube waves are clearly attenuating.}
\label{fig:image_perm}
\end{figure}

Next, we introduce the effect of porous formations in the modeled response (Figure \ref{fig:image_perm}). We interpret the location of the porous layers (Table \ref{table:PZ_synth}) based on earlier studies \cite{Kiguchi2001, Kiguchi2003SEGJ} and Table \ref{table:PZ}. However, it was necessary to change slightly the locations of a few permeable structures in \citeA{Kiguchi2003SEGJ}, viz. no. 7--9 in Table \ref{table:PZ}, so that they correspond to the results of the detailed core mapping of \citeA{Tanaka2001}. As mentioned earlier, \citeA{Kiguchi2003SEGJ} analyzed open fractures and porous layers separately using different models (Table \ref{table:PZ}). The theory developed in this study can be applied to porous layers (``PZ'' or no. 1, 4, 8, 9 in Table \ref{table:PZ}) because it associates porous formations with poroelastic deformation. On the other hand, open fractures (``OF'' or no. 2, 3, 5--7 in Table \ref{table:PZ}) have been earlier modeled considering dynamic motion of viscous fluid at a thin layer of water sandwiched between two elastic layers \cite<e.g.,>[]{Ionov2007,Bakku2013}. In this study, therefore, we assign the values of permeability to each layer as follows (Table \ref{table:PZ_synth}). First, the permeability of the layers analyzed using the permeable-zone (``PZ'') model in the study of \citeA{Kiguchi2003SEGJ} is assigned such that their hydraulic transmissivity (the product of permeability and porous-layer thickness) is the same as that reported in \citeA{Kiguchi2003SEGJ}. Next, the other permeable structures analyzed using the open-fracture (``OF'') model in \citeA{Kiguchi2003SEGJ} are represented by a porous layer with the minimum thickness (0.2 m) as assumed in our model. The permeability of the latter structures is determined such that the tube-wave amplitude of the calculated pressure response is of the same order of magnitude as the observed data, after repeated calculation of the response using various permeability values. We find that the permeabilities of these structures (no. 2, 3, 5--7) require one to two orders of magnitude larger than those of the other structures (no. 1, 4, 8, 9). This is mainly because of the large tube-wave amplitude in the observed data and the small thickness (0.2 m) assumed in the model.

 \begin{table}
 \caption{Location and value of permeability of porous formations assumed in calculating the pressure response. }
 \centering
 \begin{tabular}{crc}
 \hline
   No.  & Depth (m) & Permeability (darcy)  \\
 \hline
   1  & 181.0 -- 182.0 & 0.2  \\
   2  & 232.0 -- 232.2 & 20  \\
   3  & 247.0 -- 247.2 & 10  \\
   4  & 290.0 -- 291.0 & 0.63  \\
   5  & 427.0 -- 427.2 & 5  \\
   6  & 471.0 -- 471.2 & 5  \\
   7  & 523.2 -- 523.6 & 1   \\
   8  & 623.0 -- 625.2 & 1.8  \\
   9  & 668.0 -- 672.8 & 0.8  \\
 \hline
 \end{tabular}
\label{table:PZ_synth}
 \end{table}

At the depth with non-zero permeability, the porosity is obtained from the downhole logging data (Figure \ref{fig:intro_data}b). In addition to porosity, the Skempton coefficient ($B$) is necessary, which is essential to describe the efficiency of transmitting the elastic stress into the pore pressure (equation \ref{eq:pext}). However, obtaining the Skempton coefficient requires at least one additional independent material property, e.g., the grain bulk modulus (see equation \ref{eq:B}). The Skempton coefficient is almost 1 for soft materials such as soils and 0.1--0.3 for hard rocks. In this study, we assume $B=1$, which enables us to discuss the value of maximum pressure due to fluid infiltration at a layer with a given permeability. 

The modeled pressure response with porous layers present at the nine depth locations (Figure \ref{fig:image_perm}a) illustrates that the tube waves are additionally generated at these locations. Furthermore, the propagating tube waves experience reflections and transmissions (attenuation) in these porous layers. We also investigate the effects of background permeability (uniform permeability at all layers except for the nine depth locations). Figure \ref{fig:image_perm}(b) shows the pressure response assuming the background permeability to be 0.01 darcy in the fault damage zone (the depth larger than 426 m). Due to a continuous distribution of non-zero permeability, the generated tube waves show larger attenuation (see the red ellipse in Figure \ref{fig:image_perm}b), which is also visible in the field data (Figure \ref{fig:image_elastic}a)

\begin{figure}
\centering
 \noindent\includegraphics[width=\textwidth]{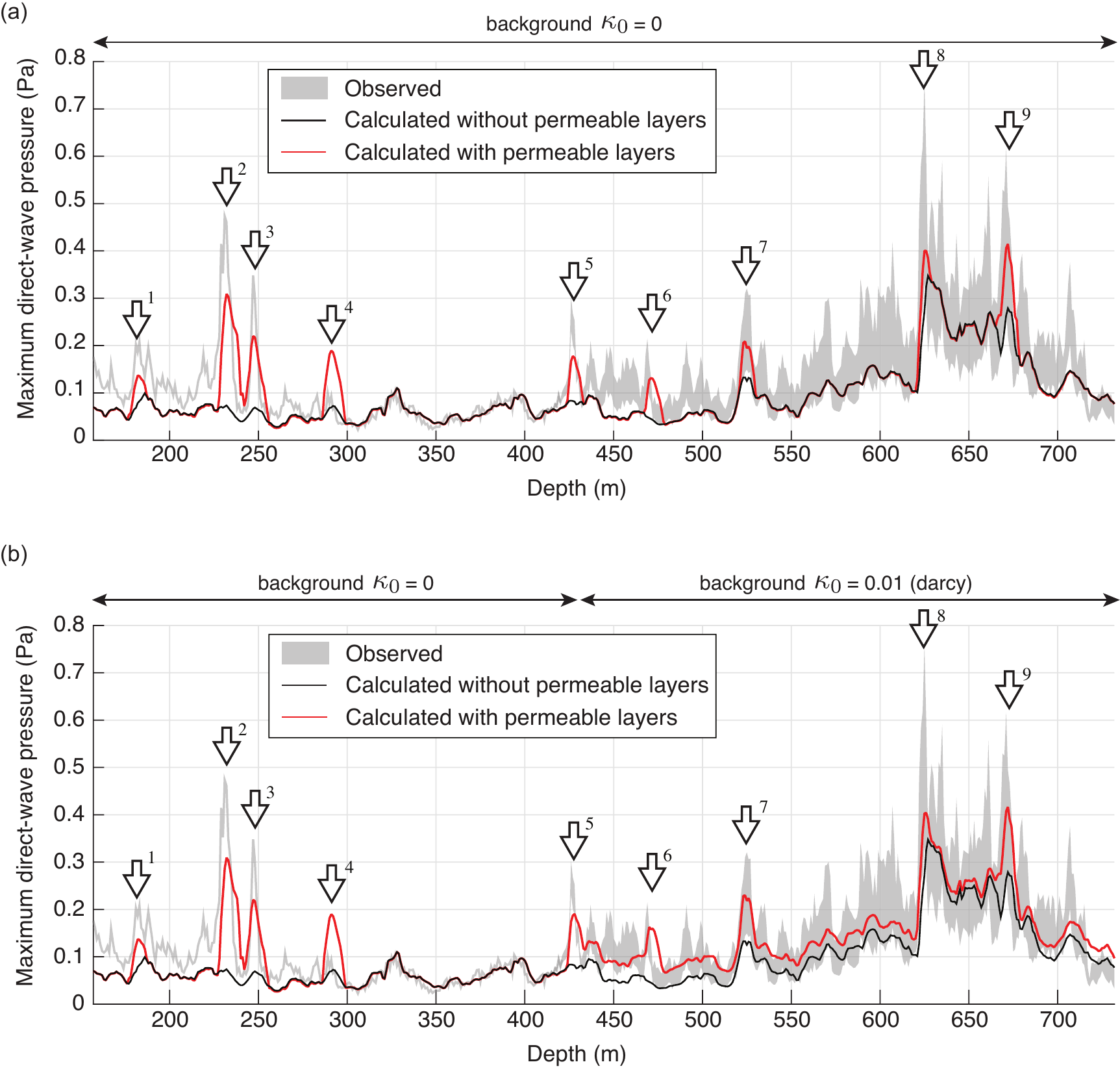}
\caption{(a) Maximum pressure amplitudes in the direct-wave waveform using the calculated response without presence of any porous formation (black line, Figure \ref{fig:image_elastic}c), and those assuming porous formations to be present at the nine depth locations (red line, Figure \ref{fig:image_perm}a). The amplitude of the field (hydrophone) data is also shown after normalization using the calculated response at 300--310 m depth (see main text). (b) Same as (a), but the red line shows the effect of additional fluid flow due to non-zero background permeability below 426 m depth (Figure \ref{fig:image_perm}b).}
\label{fig:amp}
\end{figure}

Finally, we check quantitatively the amplitude of the pressure response. The maximum amplitude of the direct wave in the modeled response (Figure \ref{fig:image_elastic}c and Figure \ref{fig:image_perm}a) is calculated and compared with the field data (Figure \ref{fig:amp}a). Since the absolute amplitude is unknown, we calibrate the field data using the maximum amplitude observed between 300 m and 310 m depths. A scaling factor is derived so that the average amplitude corresponds with the modeled data, assuming that porous formations are absent in this depth range (300--310 m). The amplitude of the modeled response (black line in Figure \ref{fig:amp}a) generally agrees with the field data. Furthermore, the inclusion of porous layers at the nine depth locations (Table \ref{table:PZ_synth}) simulates better the local increase in the amplitude as observed in hydrophone data (red line in Figure \ref{fig:amp}a). Apparently, the non-zero background permeability at the fault damage zone is responsible for a uniform increase in the amplitude (red line in Figure \ref{fig:amp}b), which agrees well with the observation. 

\section{Discussion}
The synthetic modeling experiments show that the three mechanisms, viz. elastic impedance boundary, porous formation, and borehole irregularities, contribute to the amplitudes of tube waves observed at a fault zone. The amplitude ratio between the tube wave and the incident P wave (tube-to-P-wave amplitude ratio) has been used in the past to estimate hydraulic properties \cite{Beydoun1985, Hardin1987, Li1994, Kiguchi2001}, assuming that the tube waves are generated only due to the presence of permeable zones. Furthermore, these prevalent approaches isolate first a tube wave from an incident P wave using a seismic record from a receiver which is located at a few wavelengths away from the depth where the tube wave is generated. This processing requires an assumption that the P wave and the tube wave are invariant within these depths. Our results suggest that such simple assumptions may not be appropriate in analyzing data at a fault zone where seismic velocities and density changes considerably. For example, the tube wave generated at the main shear zone (MSZ) of the Nojima fault contains the effect of the significant elastic impedance boundary at MSZ as well as that of the porous formations (Figure \ref{fig:image_elastic}c). The borehole irregularities at the heavily deformed zone as MSZ introduces additional effects in the amplitude (Figure \ref{fig:image_elastic}d). The theory and the modeling approach developed in this study enable us to calculate the pressure response that includes all these mechanisms. Moreover, the time to compute the waveforms is significantly shorter when the new approach is used. For example, calculating the synthetic data for a three-layer model (Section \ref{Sec:TPL}) takes a few seconds using the simplified theory implemented in MATLAB, whereas that using the poroelastic FD takes $\sim$15 minutes (the FD code is implemented in Julia). Using the simplified theory, calculating the field hydrophone response consisting of 2876 layers takes only about 30 seconds. Therefore, the modeling approach proposed in this study is valuable for developing new inversion methods that estimate material (hydraulic) properties by optimizing observed waveforms without the too simple assumptions that are made so far. 

The field hydrophone data at the fault zone suggests a relatively large attenuation of P waves ($Q_P$ = 17--25, Section \ref{Sec:Field_Preproc}). The quality factors estimated earlier for this site are $Q_S$ = 25--90 based on earthquake observations \cite{Li1998,Kuwahara1999,Mizuno2006}. Direct P waves in hydrophone VSP data generally exhibit a lower signal-to-noise ratio than geophone data due to interferences of tube waves \cite{Krohn1992}. Therefore, we evaluate the errors in the estimated quality factor by applying the same attenuation analyses to our modeled hydrophone response. We find that the apparent frequency shifts due to tube-wave interference do not explain the long-wavelength trend in the frequency shifts in the observed data (\ref{Append:Source_Q}). This result suggests that the estimated attenuation represents a true feature in this field. The consistency of the estimated $Q$ corroborates this finding; the amplitude measured in the field hydrophone data after inverse $Q$ filtering, where $Q$ is estimated from the frequency components and not from the amplitudes, matches well with that in the synthetic data (Section \ref{Sec:Field_Synth}). 

There are several possibilities to explain the observed P-wave attenuation. First of all, the scattering attenuation due to seismic velocity fluctuations in the spatial scale of the acoustic log (0.2 m spacing) is excluded because our modeled response is based on those velocities. Therefore, possible attenuation mechanisms are scattering due to a smaller-scale fluctuation in elasticity such as cracks \cite{Hudson1981} and anelasticity due to grain-scale/mesoscopic fluid flow \cite<e.g.,>[]{Dvorkin1994,Chapman2003}. The field data suggest different $Q$ values at the host rock region (157 m -- 426 m depth) and at the fault damage zone (426 m -- 732 m depth). The core analyses indicate a clear difference in microstructures and chemical compositions between these two regions \cite{Tanaka2001}. \citeA{Ito2005} analyze the formation microimager (FMI) resistivity log and shear-wave anisotropy using DSI. They show that mesoscopic fractures exist throughout the survey depth. These authors further illustrate significant differences in fracture distributions between the host rock region and the fault damage zone, specifically in terms of the orientation and spacing of the fractures and resulting shear-wave anisotropy. Therefore, the presence of microcracks and mesoscopic fractures and the changes in their geometrical distribution can potentially explain the observed attenuation. 

We assume the sonic velocities to be frequency independent in synthetic modeling of field hydrophone data. At the same time, we consider those velocities at the porous layers to represent the low-frequency limit based on the Gassmann equation. The $V_P$ is estimated from monopole sonic waveform with a center frequency of 10 kHz and $V_S$ from dipole sonic waveform with a center frequency of 2 kHz \cite{Ito2005}. Using the porosity log from the field (Figure \ref{fig:intro_data}b), the average value of the Biot characteristic frequency \cite<$f_c=\phi\eta/2\pi\rho_{\rm f}\kappa_0$,>[]{MavkoCh6} is approximately 1 MHz for $\kappa_0=0.01$ darcy and 10 kHz for $\kappa_0=1$ darcy. Therefore, the assumption that the velocity log data represents the low-frequency regime is justified. Note, however, that anelastic attenuation due to grain-scale/mesoscopic fluid flow \cite<e.g.,>[]{Dvorkin1994,Chapman2003} can show the characteristic frequency much lower than that from the Biot theory \cite{MavkoCh6}. Hence, the sonic velocities at the fault damage zone should be used carefully, where the attenuation is estimated to be large. In this vein, incorporating seismic attenuation in our forward modeling approach is possible through introducing the attenuation into the elastic wave potential field. Such extension and the effect of velocity dispersion on borehole pressure response require further research. Admittedly, the existing conventional approaches that analyze the tube-to-P-wave amplitude ratio \cite{Beydoun1985, Hardin1987, Li1994, Kiguchi2001} will suffer less from seismic attenuation. This feature is because the amplitude ratio will mitigate the effect of attenuation for direct P waves.

In this study, we exclude a tube-wave generation mechanism due to open fractures. These fractures have been modeled as a thin layer of viscous fluid sandwiched between two elastic layers \cite{Ionov2007,Bakku2013,Minato2017JGR}. Such open fractures exist throughout the survey depth in the field example shown in this paper. The model developed in this study can calculate the pressure response of fractured media, provided that porous formations can approximate them using appropriate porosity and permeability. In this regard, \citeA{Minato2017SEG} attempted to relate the parameters of a thin porous layer model with those of the open fracture model, assuming the presence of multiple fractures. Note, however, that the models discussed in \citeA{Minato2017SEG} are based on the effective-source formulation. We show in this study that the absence of a localized difference in the tube-wave velocity in the effective-source model results in a large deviation from the Biot theory (Section \ref{Sec:TPL}). Therefore, it is still an open question how one can incorporate accurately the tube waves at open fractures into the simplified theory, especially when the fractures are located in a heterogeneous elastic background and the borehole has irregularities (i.e., borehole radius is not constant in depth). 
 
We test a continuous permeability profile or a non-zero background permeability at the fault damage zone below 426 m depth (Section \ref{Sec:Field_Synth}). The modeled data matches well with the field data in terms of attenuation of tube waves (Figure \ref{fig:image_perm}b) and amplitude distribution (Figure \ref{fig:amp}b). The magnitude of the background permeability ($\kappa_0=0.01$ darcy) is representative of the controlled laboratory experiments on field outcrop samples (fractured granite and granite breccia) at an effective pressure that corresponds to 400--700 m depth or 7--10 MPa \cite<see Fig. 7 in>[]{Mizoguchi2008}. However, this does not exclude the possibility that the actual permeability distribution is highly localized at a finite number of thin porous layers or open fractures. We expect that such a discrete permeability profile would produce a similar tube-wave attenuation and amplitude distribution as in the case when a continuous permeability profile is assumed. This similarity is due to the finite-frequency nature of the wavefield measurements.

\section{Conclusion}
Hydrophone VSP has the potential for the efficient monitoring of hydraulic permeability using a borehole. To better understand the dynamic interaction between the borehole fluid and porous formations, we have presented a theory that enables calculation of the pressure response in an irregular-radius borehole embedded in layered poroelastic media when a plane P wave is traveling along the borehole. We solve the system of equations using a propagator matrix formulation. We derive the analytical closed-form expressions for pressure amplitudes, considering three mechanisms that affect the generation of tube waves: an elastic impedance boundary, a thin poroelastic layer sandwiched between two elastic layers, and a step-like change in the borehole radius. The pressure waveforms calculated using the newly developed theory show a good agreement with those from fully coupled (acoustic-elastic-poroelastic) numerical solutions of Biot poroelasticity equations. We illustrate that the amplitude of the upgoing and the downgoing tube waves produced by the elastic impedance contrast has opposite polarities. In the case of the tube wave generated at a thin porous layer, the amplitude is frequency-dependent. Furthermore, the upgoing and the downgoing tube wave have different waveforms. Comparing the tube wave amplitude derived in this research (assuming the presence of a thin porous layer) with that derived in the past studies (assuming effective source volume), we find that the effective-source models of the past are not consistent with the Biot solutions. This inconsistency is mainly because these models do not consider the effect of the local change in the tube-wave velocity in the porous layer. 

We have discussed the role of the three mechanisms in a field VSP dataset acquired at an active fault zone in Japan (Nojima fault). We calculate the pressure response using the downhole logging data for seismic velocities, density, porosity, and borehole radius. We find that the tube waves generated due to elastic impedance boundaries, especially those around the main shear zone of the fault, have large amplitudes. We also find that tube waves are generated with non-negligible amplitudes at depths where the borehole radius increases at the main shear zone. Furthermore, the presence of porous layers explains the local increase in the pressure in the field hydrophone data. The analytical solutions and the modeling approach developed in this study enable fast and accurate computation of the pressure response, taking into account all three mechanisms. VSP using the new model will be crucial in monitoring hydraulic properties at a fault zone by accurate interpretation of tube waves.

\newpage

\appendix

\section{Governing equations of the borehole-coupling theory}
\label{Append:Ionov}
In this appendix, we briefly review the theory of \citeA{Ionov1996} in order to derive equation \ref{eq:PDE1} in the main text. We assume low-frequency, small-amplitude, azimuthally symmetric wave motion of a fluid in a circular cylinder (i.e., a borehole). In this case, pressure and vertical particle velocity are described by the following partial differential equation:
\begin{linenomath*}
 \begin{equation}
  \frac{\partial}{\partial z}
   \begin{pmatrix}
    p \\ 
    v_z
   \end{pmatrix}=
  i\omega\begin{pmatrix}
	  0 & \rho_f \\
	  K_{\rm f}^{-1} & 0
	 \end{pmatrix}
         \begin{pmatrix}
	  p \\ 
	  v_z
	 \end{pmatrix}+
	 \begin{pmatrix}
	  0 \\ 
	  -\frac{2}{R}\left.v_r\right\vert_{r=R}
	 \end{pmatrix},
	 \label{eq:PDE0}
 \end{equation}
\end{linenomath*}
where the second term on the right-hand side is the contribution of the radial motion of the fluid at the borehole wall ($r=R$). The radial motion is expressed as
 \begin{linenomath*}
 \begin{equation}
  \left.v_r\right\vert_{r=R}=R\left(\frac{-i\omega p}{2\mu}+\frac{-i\omega\sigma^{\rm ext}_{\rm eff}}{E}\right)+v_{\rm ft}.
   \label{eq:vr}
 \end{equation}
\end{linenomath*}
In equation \ref{eq:vr}, the first term indicates quasi-static elastic deformation, and the second term indicates fluid infiltration due to the presence of a porous formation \cite{Ionov1996}.  Within the porous formation, the pore pressure $p_{\rm por}(r,\omega)$ satisfies the following diffusion equation:
 \begin{linenomath*}
 \begin{equation}
  -i\omega p_{\rm por}=a^2\nabla^2_r p_{\rm por},
\label{eq:PDE_por}
 \end{equation}
\end{linenomath*}
where $\nabla^2_r=\partial^2/\partial r + r^{-1}\partial/\partial r$ indicates the radial part of the Laplacian operator, and the pressure diffusivity $a$ is related to the properties of the porous formation:
 \begin{linenomath*}
 \begin{equation}
  a^2=\frac{\kappa_0 K_f}{\eta\phi},
 \end{equation}
\end{linenomath*}
where $\kappa_0$ is the static permeability, $\eta$ the dynamic viscosity, and $\phi$ the porosity. Equation \ref{eq:PDE_por} is derived from Darcy's law and the continuity equation, followed by a linearization \cite{Ionov1996}. A non-divergent solution of equation \ref{eq:PDE_por} is
 \begin{linenomath*}
 \begin{equation}
  p_{\rm por}(r,\omega)=\left(p-p^{\rm ext}_{\rm por}\right)\frac{K_0\left(\sqrt{-i\omega}r/a\right)}{K_0\left(\sqrt{-i\omega}R/a\right)}+p^{\rm ext}_{\rm por},
 \label{eq:p_por}
 \end{equation}
\end{linenomath*}
where $K_0$ is the modified Bessel function of the second kind of order zero, the pressure at the borehole $p=p_{\rm por}(R,\omega)$, and the pressure at a large distance from the borehole $p^{\rm ext}_{\rm por}=\lim_{r\to\infty}p_{\rm por}(r,\omega)$. Finally, the infiltration velocity ($v_{\rm ft}$ in equation \ref{eq:vr}) can be obtained by the following relation:
 \begin{linenomath*}
 \begin{align}
  v_{\rm ft}&=-\frac{\kappa_0}{\eta}\left.\frac{\partial p_{\rm por}}{\partial r}\right\vert_{r=R} \\
            &=-i\omega R\frac{\phi}{K_f}\left(p-p^{\rm ext}_{\rm por}\right)\Phi\left(\sqrt{-i\omega t_f}\right),
 \label{eq:vft}
 \end{align}
\end{linenomath*}
where $\Phi(w)=w^{-1}K_1(w)/K_0(w)$ and $t_f=R^2\phi\eta/\kappa_0K_f$. Finally, equation \ref{eq:PDE1} in the main text can be derived using equations \ref{eq:PDE0}, \ref{eq:vr}, and \ref{eq:vft}.

\section{Potential amplitudes of a normally incident plane P wave at a stack of layers}
\label{Append:phiE}
In this study, we consider the displacement potential of the external elastic waves in the form of $\phi_{\rm E}(z)=D_{\rm E}\exp(ik_pz)+U_{\rm E}\exp(-ik_pz)$. Calculating the potential amplitudes in a stack of elastic layers due to a normally incident plane P wave is trivial, and one can find the solution elsewhere. In order to replicate the analytical tube-wave amplitude derived in this study (Section \ref{Sec:Simple_Config}), we briefly summarize the solution in this appendix, using the same notation as in the main text.

First, the vertical velocity and the normal stress are derived from the potential as, $v_z^E=-i\omega\partial \phi_{\rm E}/\partial z$ and $\sigma_{zz}=-\rho\omega^2\phi_{\rm E}$, respectively. Considering the continuity of $v_z^E$ and $\sigma_{zz}$ at the $n$-th boundary, the potential-amplitude vector of the plane P wave,  
$\mathbf{u}^E_n=(U_{\rm E}^{(n)},D_{\rm E}^{(n)})^{\rm T}$, obeys the following relation:
\begin{linenomath*}
 \begin{equation}
 \mathbf{u}_n^E=\mathbf{M}^E_n(z_n)\mathbf{u}^E_{n+1},
  \label{eq:unE}
 \end{equation}
\end{linenomath*}
 where the matrix $\mathbf{M}_n^E$ is defined as 
\begin{linenomath*}
 \begin{equation}
 \mathbf{M}^E_n(z)=\frac{1}{2}
   \begin{pmatrix}
    a^E_1 e^{i(k_{p_n}-k_{p_{n+1}})z} && a^E_2 e^{i(k_{p_n}+k_{p_{n+1}})z} \\ 
    a^E_2 e^{-i(k_{p_n}-k_{p_{n+1}})z} && a^E_1 e^{-i(k_{p_n}+k_{p_{n+1}})z} 
   \end{pmatrix},
    \label{eq:MnE}
 \end{equation}
\end{linenomath*}
\begin{linenomath*}
 \begin{align}
 &a^E_1=\frac{\rho_{n+1}}{\rho_n}+\frac{k_{p_{n+1}}}{k_{p_n}}, \\
 &a^E_2=\frac{\rho_{n+1}}{\rho_n}-\frac{k_{p_{n+1}}}{k_{p_n}}.
 \end{align}
\end{linenomath*}
Successively applying equation \ref{eq:unE}, we obtain the following relation:
\begin{linenomath*}
 \begin{align}
 \mathbf{u}^E_1&=\prod_{i=1}^{N-1}\mathbf{M}^E_i(z_i)\mathbf{u}^E_{N} \nonumber \\
  &=\mathbf{M}_T^E\mathbf{u}_N^E.
  \label{eq:unu1E} 
 \end{align}
\end{linenomath*}
We consider a radiation condition for an infinite half-space at the top and the bottom layers. The condition can be represented as $\mathbf{u}^E_1=(U_{\rm E}^{(1)},D_{\rm E}^{(1)})^{\rm T}$ and $\mathbf{u}^E_N=(0,D_{\rm E}^{(N)})^{\rm T}$. We solve equation \ref{eq:unu1E} for $U_{\rm E}^{(1)}$ and $D_{\rm E}^{(N)}$ with a known value for $D_{\rm E}^{(1)}$ (incident wave). The amplitudes for all layers can be obtained from equation \ref{eq:unE}. 

The reflection and transmission coefficients ($R_{\rm E}$, $T_{\rm E}$) for a downgoing incident wave are derived considering $N=2$, $\mathbf{u}^E_1=(R_{\rm E},1)^{\rm T}$, and $\mathbf{u}^E_2=(0,T_{\rm E})^{\rm T}$:
\begin{linenomath*}
 \begin{align}
 R_{\rm E}&= \frac{\rho_1k_{p_2}-\rho_2k_{p_1}}{{\rho_1k_{p_2}+\rho_2k_{p_1}}}\nonumber \\
 T_{\rm E}&= \frac{2\rho_1k_{p_2}}{{\rho_1k_{p_2}+\rho_2k_{p_1}}}.
  \label{eq:RTcoef_E} 
 \end{align}
\end{linenomath*}

\section{Calculation of discontinuities in pressure and velocity due to continuous sources by the propagator matrix method}
\label{Append:delta}
The discontinuities defined in the boundary condition (equations \ref{eq:BC1}, \ref{eq:BC2}) include contributions due to the continuous sources located over depth within a layer. In this appendix, the discontinuities (equations \ref{eq:delta_pE}, \ref{eq:delta_pft}) are derived using the propagator matrix method. The governing equation (equation \ref{eq:PDE1}) can be written as,
\begin{linenomath*}
 \begin{equation}
 \frac{d\mathbf{f}(z)}{d z}=\mathbf{A}(z)\mathbf{f}(z)+\mathbf{g}(z),
  \label{eq:dfdz} 
 \end{equation}
\end{linenomath*}
where $\mathbf{f}=(p,v_z)^{\rm T}$, and $\mathbf{g}$ represents the source term. The solution to equation \ref{eq:dfdz} may be written as \cite{Aki2002_Ch7.2.2},
\begin{linenomath*}
 \begin{equation}
 \mathbf{f}(z)=\int_{z_0}^z\mathbf{P}(z,\zeta)\mathbf{g}(\zeta)d\zeta+\mathbf{P}(z,z_0)\mathbf{f}(z_0),
  \label{eq:f} 
 \end{equation}
\end{linenomath*}
where $\mathbf{P}$ is the propagator matrix.
\begin{linenomath*}
 \begin{equation}
 \mathbf{P}(z,z_0)=\frac{1}{2}
   \begin{pmatrix}
    e^{ik(z-z_0)}+e^{-ik(z-z_0)} && \rho_f C_T\left(e^{ik(z-z_0)}-e^{-ik(z-z_0)}\right) \\ 
    \frac{1}{\rho_f C_T}\left(e^{ik(z-z_0)}-e^{-ik(z-z_0)}\right) && e^{ik(z-z_0)}+e^{-ik(z-z_0)} 
   \end{pmatrix}.
    \label{eq:P}
 \end{equation}
\end{linenomath*}
Equation \ref{eq:f} corresponds to equations \ref{eq:p} and \ref{eq:vz}. The first term of equation \ref{eq:f} is the discontinuities in pressure and velocity ($\Delta p$ and $\Delta v_z$). The discontinuities can be calculated analytically using equations \ref{eq:f}, \ref{eq:P}, \ref{eq:sext}, and \ref{eq:pext}. In equations \ref{eq:f} and \ref{eq:P}, we can consider the contributions separately from the source term related to the elastic deformation (the term including $\sigma_{\rm eff}^{\rm ext}$ in $\mathbf{g}$) and the fluid infiltration (the term including $p_{\rm por}^{\rm ext}$ in $\mathbf{g}$). In this case, the discontinuities can be represented by equations \ref{eq:delta_pE} to \ref{eq:delta_vft} at the $n$-th layer ($z_0=z_{n-1}$). The functions $I_1$--$I_4$ in these equations are written as
\begin{linenomath*}
 \begin{align}
 I_1(z_{n-1},z)=&\frac{e^{i k_n z}}{i(-k_n+k_{p_n})}\{e^{i\left(-k_n+k_{p_n}\right)z}-e^{i\left(-k_n+k_{p_n}\right)z_{n-1}}\} \nonumber \\
  &-\frac{e^{-i k_n z}}{i(k_n+k_{p_n})}\{e^{i\left(k_n+k_{p_n}\right)z}-e^{i\left(k_n+k_{p_n}\right)z_{n-1}}\}, \label{eq:I1} \\
 I_2(z_{n-1},z)=&\frac{e^{i k_n z}}{i(-k_n-k_{p_n})}\{e^{i\left(-k_n-k_{p_n}\right)z}-e^{i\left(-k_n-k_{p_n}\right)z_{n-1}}\} \nonumber \\
  &-\frac{e^{-i k_n z}}{i(k_n-k_{p_n})}\{e^{i\left(k_n-k_{p_n}\right)z}-e^{i\left(k_n-k_{p_n}\right)z_{n-1}}\}, \label{eq:I2} \\
 I_3(z_{n-1},z)=&\frac{e^{i k_n z}}{i(-k_n+k_{p_n})}\{e^{i\left(-k_n+k_{p_n}\right)z}-e^{i\left(-k_n+k_{p_n}\right)z_{n-1}}\} \nonumber \\
  &+\frac{e^{-i k_n z}}{i(k_n+k_{p_n})}\{e^{i\left(k_n+k_{p_n}\right)z}-e^{i\left(k_n+k_{p_n}\right)z_{n-1}}\}, \label{eq:I3} \\
 I_4(z_{n-1},z)=&\frac{e^{i k_n z}}{i(-k_n-k_{p_n})}\{e^{i\left(-k_n-k_{p_n}\right)z}-e^{i\left(-k_n-k_{p_n}\right)z_{n-1}}\} \nonumber \\
  &+\frac{e^{-i k_n z}}{i(k_n-k_{p_n})}\{e^{i\left(k_n-k_{p_n}\right)z}-e^{i\left(k_n-k_{p_n}\right)z_{n-1}}\}. \label{eq:I4}
 \end{align}
\end{linenomath*}

In the special case where we consider the contribution from infinity (i.e., $z_{n-1}=-\infty$), equations \ref{eq:I1} to \ref{eq:I4} can be written as 
\begin{linenomath*}
 \begin{align}
 I_1(-\infty,z)&=\frac{2k}{i\left(k_{p}^2-k^2\right)}e^{i k_{p}z}, \label{eq:I1inf} \\
 I_2(-\infty,z)&=\frac{2k}{i\left(k_{p}^2-k^2\right)}e^{-i k_{p}z},  \label{eq:I2inf} \\
 I_3(-\infty,z)&=\frac{2k_{p}}{i\left(k_{p}^2-k^2\right)}e^{i k_{p}z},  \label{eq:I3inf} \\
 I_4(-\infty,z)&=\frac{-2k_{p}}{i\left(k_{p}^2-k^2\right)}e^{-i k_{p}z},   \label{eq:I4inf}
 \end{align}
\end{linenomath*}
where we ignore the terms associated with the infinite delay time \cite{White1953}.

\section{Tube-wave potential amplitudes in homogeneous elastic media}
\label{Append:homo}
In this appendix, we derive the potential amplitudes of the borehole fluid $(U_{\rm f},D_{\rm f})$ when a downgoing P wave propagates in homogeneous elastic media. The results are then utilized for the radiation condition at half-spaces in the propagator matrix formulation (Section \ref{Sec:Elastic}). 

\begin{figure}
\centering
 \noindent\includegraphics{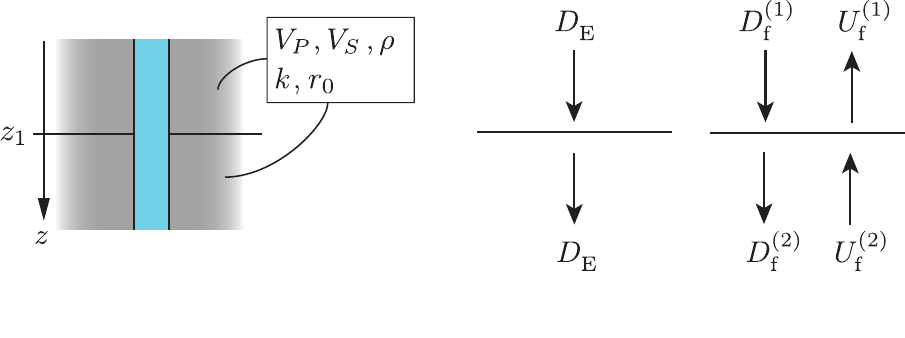}
\caption{A borehole with a constant radius ($r_0$) embedded in a homogeneous elastic medium. The elastic-wave potential amplitude contains the downgoing incident wave only ($D_{\rm E}$). The tube-wave potential amplitudes contain upgoing and downgoing waves at each layer ($U^{(i)}_{\rm f}$ and $D^{(i)}_{\rm f}$).}
\label{fig:Geom_Elastic_homo}
\end{figure}

We consider the boundary located at $z=z_1$ in the homogeneous elastic medium (see Figure \ref{fig:Geom_Elastic_homo}). In this case, equation \ref{eq:unu1} can be written as
\begin{linenomath*}
 \begin{align}
 \mathbf{u}_1&=\mathbf{M}_1(z_1)\mathbf{u}_{2} \nonumber \\
             &=\mathbf{u}_{2},
  \label{eq:un_homo}
 \end{align}
\end{linenomath*}
where we used the relation $k_1=k_2=k$ and $\mathbf{M}_1=\mathbf{I}$. The absence of the source vector $\mathbf{S}_T$ in equation \ref{eq:un_homo} is due to the single boundary and the absence of the radius change (i.e., $\Delta v_{\rm q}=0$). 

We define $\mathbf{u}_1$ such that the pressure and particle velocity at $z=z_1$ are equivalent to the discontinuity due to the continuous source at the infinite half-space above $z_1$:
\begin{linenomath*}
 \begin{equation}
  \left.
   \begin{pmatrix}
    p \\
    v_z
   \end{pmatrix}\right|_{z=z_1}:=
   \begin{pmatrix}
    \Delta p_{\rm E}^{(1)}(-\infty,z_1) \\ 
    \Delta v_{\rm E}^{(1)}(-\infty,z_1)
   \end{pmatrix}.
  \label{eq:p_and_dp}
 \end{equation}
\end{linenomath*}
The discontinuities $\Delta p_{\rm E}^{(1)}$ and $\Delta v_{\rm E}^{(1)}$ in equation \ref{eq:p_and_dp} contain the contribution of the source continuously located between $-\infty$ and $z_1$ or the functions $I_i(-\infty,z_1)$ where $i=1-4$ in equations \ref{eq:delta_pE} and \ref{eq:delta_vE}. The values of $I_i(-\infty,z_1)$ can be analytically calculated using equations \ref{eq:I1inf} to \ref{eq:I4inf}.

The explicit solutions of $U_{\rm f}^{(1)}$ and $D_{\rm f}^{(1)}$ can be obtained from equation \ref{eq:p_and_dp} by replacing $p$ and $v_z$ by the potential amplitudes, substituting equations \ref{eq:delta_pE}, \ref{eq:delta_vE}, \ref{eq:I1inf}, and \ref{eq:I3inf}, and considering the presence of a downgoing elastic wave ($D_{\rm E}^{(1)}=D_{\rm E}^{(2)}=D_{\rm E}$ and $U_{\rm E}^{(1)}=U_{\rm E}^{(2)}=0$): 
\begin{linenomath*}
 \begin{align}
    U_{\rm f}^{(1)} &= \frac{e^{ikz_1}}{2\rho_f\omega^2k}
    \{k\Delta p_{\rm E}^{(1)}(-\infty,z_1)-\rho_f\omega\Delta v_{\rm E}^{(1)}(-\infty,z_1)\}   \nonumber \\
                    &=\frac{k_{p}}{k\left(k+k_{p}\right)}A_PD_{\rm E} e^{i(k+k_p)z_1}, \label{eq:Uf1} \\
    D_{\rm f}^{(1)} &= \frac{e^{-ikz_1}}{2\rho_f\omega^2k}
    \{k\Delta p_{\rm E}^{(1)}(-\infty,z_1)+\rho_f\omega\Delta v_{\rm E}^{(1)}(-\infty,z_1)\},  \nonumber \\
                    &=\frac{k_{p}}{k\left(k-k_{p}\right)}A_PD_{\rm E} e^{-i(k-k_p)z_1}. \label{eq:Df1} 
 \end{align}
\end{linenomath*}

Next, we show that the above solution (equations \ref{eq:Uf1} and \ref{eq:Df1}) indicates the borehole response of the downgoing plane P wave. To this end, we consider the pressure response at $z=z_1\pm Z$ where $Z > 0$. From equation \ref{eq:p}, we obtain the pressure at the medium above the boundary ($z=z_1-Z$):
\begin{linenomath*}
 \begin{align}
p(z_1 - Z)&=\rho_f\omega^2\left(D^{(1)}_{\rm f} e^{ik(z_1 - Z)}+U^{(1)}_{\rm f} e^{-ik(z_1 - Z)}\right)+\Delta p^{(1)}_{\rm E}(z_1,z_1 - Z) \nonumber \\
  &=-\rho_f C_T \omega\frac{2 k k_p}{k_p^2-k^2}A_P D_{\rm E} e^{i k_p (z_1 - L)}, \label{eq:phomo1}
 \end{align}
\end{linenomath*}
and at the medium below the boundary ($z=z_1+Z$):
\begin{linenomath*}
 \begin{align}
p(z_1 + Z)&=\rho_f\omega^2\left(D^{(2)}_{\rm f} e^{ik(z_1 + Z)}+U^{(2)}_{\rm f} e^{-ik(z_1 + Z)}\right)+\Delta p^{(2)}_{\rm E}(z_1,z_1 + Z) \nonumber \\
  &=-\rho_f C_T \omega\frac{2 k k_p}{k_p^2-k^2}A_P D_{\rm E} e^{i k_p (z_1 + L)}, \label{eq:phomo2}
 \end{align}
\end{linenomath*}
where we use the relation $U^{(2)}_{\rm f}=U^{(1)}_{\rm f}$ and $D^{(2)}_{\rm f}=D^{(1)}_{\rm f}$ (equation \ref{eq:un_homo}).
These equations indicate that, due to the interaction among the tube waves propagating in the opposite directions (upgoing and downgoing) and the source continuously located over depth, only one downgoing wave having P-wave velocity remains. The pressure amplitude of the P wave in equations \ref{eq:phomo1} and \ref{eq:phomo2} is equivalent to the known analytical solution of the borehole response due to a plane P wave using quasi-static approximation \cite<equation 5.36 in>[]{White1983} and that using the exact boundary conditions at low frequencies \cite<equations 5 and C-5 in>[]{Schoenberg1986}.

\section{Biot theory of linear poroelasticity in a borehole environment}
\label{Append:Biot}
In this research, we verify the developed theory with the numerical solutions to the fully-coupled (acoustic-elastic-poroelastic) problem using the finite-difference (FD) method in the cylindrical coordinate system. The governing equations in the FD method are based on Biot dynamic poroelasticity \cite{Biot1956a,Biot1956b,Biot1962}. The stress-strain relation in the cylindrical coordinate system can be written as \cite<e.g.,>[]{Sidler2014GJI,Ou2019GJI}:
\begin{linenomath*}
 \begin{align}
  \dot{p_f}&=-C\left(\frac{v_r^{(u)}}{r}+\frac{\partial v_r^{(u)}}{\partial r}+\frac{\partial v_z^{(u)}}{\partial z}\right)-M\left(\frac{v_r^{(w)}}{r}+\frac{\partial v_r^{(w)}}{\partial r}+\frac{\partial v_z^{(w)}}{\partial z}\right), \label{eq:pf} \\
  \dot{\tau_{rz}}&=\mu\left(\frac{\partial v_z^{(u)}}{\partial r}+\frac{\partial v_r^{(u)}}{\partial z}\right), \\
  \dot{\tau_{rr}}&=(H-2\mu)\left(\frac{v_r^{(u)}}{r}+\frac{\partial v_z^{(u)}}{\partial z}\right)+H\frac{\partial v_r^{(u)}}{\partial r}+C\left(\frac{v_r^{(w)}}{r}+\frac{\partial v_r^{(w)}}{\partial r}+\frac{\partial v_z^{(w)}}{\partial z}\right), \\
  \dot{\tau_{zz}}&=(H-2\mu)\left(\frac{v_r^{(u)}}{r}+\frac{\partial v_r^{(u)}}{\partial r}\right)+H\frac{\partial v_z^{(u)}}{\partial z}+C\left(\frac{v_r^{(w)}}{r}+\frac{\partial v_r^{(w)}}{\partial r}+\frac{\partial v_z^{(w)}}{\partial z}\right), \\
  \dot{\tau_{\theta\theta}}&=(H-2\mu)\left(\frac{\partial v_z^{(u)}}{\partial z}+\frac{\partial v_r^{(u)}}{\partial r}\right)+H\frac{v_r^{(u)}}{r}+C\left(\frac{v_r^{(w)}}{r}+\frac{\partial v_r^{(w)}}{\partial r}+\frac{\partial v_z^{(w)}}{\partial z}\right), \label{eq:tautt}
 \end{align}
\end{linenomath*}
where we asssume azimuthal symmetry. In equations \ref{eq:pf}--\ref{eq:tautt}, $p_f$ is the pore pressure, $\tau_{ij}$ is the total stress tensor, $v^{(u)}_i$ is the solid particle velocity, and $v^{(w)}_i$ is the relative fluid particle velocity. The equation of motion can be written as
\begin{linenomath*}
 \begin{align}
  \frac{\partial \tau_{rr}}{\partial r}+\frac{\partial \tau_{rz}}{\partial z}+\frac{\tau_{rr}-\tau_{\theta\theta}}{r}&=\rho\dot{v}^{(u)}_r+\rho_f\dot{v}^{(w)}_r, \\
  \frac{\partial \tau_{zz}}{\partial z}+\frac{\partial \tau_{rz}}{\partial r}+\frac{\tau_{rz}}{r}&=\rho\dot{v}^{(u)}_z+\rho_f\dot{v}^{(w)}_z, \\
 D_1\dot{v}^{(w)}_r+D_2v^{(w)}_r+\rho_f\dot{v}^{(u)}_r&=-\frac{\partial p_f}{\partial r}, \\
 D_1\dot{v}^{(w)}_z+D_2v^{(w)}_z+\rho_f\dot{v}^{(u)}_z&=-\frac{\partial p_f}{\partial z}. \label{eq:D1vz}
 \end{align}
\end{linenomath*}
The poroelastic parameters in equations \ref{eq:pf}--\ref{eq:D1vz} are:
\begin{linenomath*}
 \begin{align}
  H&=K_m+\frac{4}{3}\mu+M\alpha^2,\\
  C&=M\alpha, \\
  M&=\left(\frac{\alpha-\phi}{K_S}+\frac{\phi}{K_f}\right)^{-1}, \\
  \alpha&=1-\frac{K_m}{K_S}, \\
  D_1&=\frac{\mathcal{T}\rho_f}{\phi}, \\
  D_2&=\frac{\eta}{\kappa_0},
 \end{align}
\end{linenomath*}
where $H$, $C$, and $M$ are the porous formation moduli, $K_m$ is the frame bulk modulus, $K_S$ is the grain bulk modulus, $\mathcal{T}$ is the tortuosity factor, and $\alpha$ is the Biot-Willis constant. Our staggered-grid, finite-difference modeling approach is based on the one by \citeA{Guan2011}. In this approach, three different sub-domains are specified: acoustic (a borehole fluid), elastic, and poroelastic domains, and the same discretized equations are solved using the poroelastic properties at the limiting case in each domain. Furthermore, additional boundary conditions are considered at the acoustic-poroelastic interface (open-pore condition) and the elastic-poroelastic interface (closed-pore condition), see \citeA{Guan2011} and \citeA{Ou2019GJI} for more details. In our modeling approach, non-splitting perfectly matched layers (NPML) are implemented at model boundaries, and the boundary condition at $r=0$ is derived from the symmetry properties and l'H\^{o}pital's rule \cite{Mittet1996}. The location of the field properties in the staggered-grid cell can be found in \citeA{Guan2011}. Finally, we consider the initial condition of the FD modeling such that a normally incident plane P wave starts to propagate downwards in a borehole embedded in homogeneous elastic media. For this purpose, we assign initial values to the field quantities ($p_f,\tau_{ij},v^{(w)}_i,v^{(u)}_i$) in the borehole fluid and the surrounding elastic medium using the analytical solutions given by \citeA{Peng1994PhD}.

\section{Skempton coefficient and Gassmann's low-frequency limit for the poroelastic moduli}
\label{Append:B}
In the case of quasi-static, undrained condition where there is no fluid flux, one can derive the following relation from the stress-strain relation (equations \ref{eq:pf} to \ref{eq:tautt}):
\begin{linenomath*}
 \begin{align}
  \dot{p_f}&=-C\left(\frac{v_r^{(u)}}{r}+\frac{\partial v_r^{(u)}}{\partial r}+\frac{\partial v_z^{(u)}}{\partial z}\right), \\
  \frac{1}{3}\dot{\tau_{ii}}&=\left(H-\frac{4}{3}\mu\right)\left(\frac{v_r^{(u)}}{r}+\frac{\partial v_r^{(u)}}{\partial r}+\frac{\partial v_z^{(u)}}{\partial z}\right).
 \end{align}
\end{linenomath*}
It is then straightforward to derive the following relation:
\begin{linenomath*}
 \begin{align}
  p_f&=-\frac{1}{3}B\tau_{ii}, \label{eq:pf_App} \\
  B&=\frac{M\alpha}{K_m+M\alpha^2}, \label{eq:B}
 \end{align}
\end{linenomath*}
where $\tau_{ii}$ indicates the trace of the total stress tensor in the poroelastic formation. We assume $p_f=0$ at $t=0$. The coefficient $B$ is known as the Skempton coefficient \cite{Rice1976}. In the boundary condition of the diffusion equation in this study (equation \ref{eq:PDE_por}), we use the relation $\lim_{r\to\infty}p_{\rm por}(r,\omega)=p^{\rm ext}_{\rm por}$, where the pore-pressure gradient converges to zero, leading to the undrained condition. Therefore, we define $p^{\rm ext}_{\rm por}=-\frac{1}{3}B\sigma_{ii}$ (equation \ref{eq:pext}), where $\sigma_{ii}$ is the trace of the stress tensor in the elastic formation.

At the low-frequency limit where the relative fluid velocity is negligible, the undrained bulk modulus $K_u$ is identified by the Gassmann equation:
\begin{linenomath*}
 \begin{equation}
  K_u=K_m+M\alpha^2.
 \end{equation}
\end{linenomath*}
Consequently, when the simplified theory (Section \ref{Sec:Theory}) is applied at the medium defined by the poroelastic properties, $K$, $E$ and $\rho$ are calculated as,
\begin{linenomath*}
 \begin{align}
  K&=K_u,\\
  E&=\frac{9K_u\mu}{3K_u+\mu}, \\
  \rho&=(1-\phi)\rho_s+\phi\rho_f.
 \end{align}
\end{linenomath*}
Corresponding seismic velocities ($V_P$, $V_S$) are derived from $K$, $E$, and $\rho$ above.

\section{Effective-source formulation of the generated tube waves due to a thin porous layer}
\label{Append:Eff_Src}
Tube waves generated at the zone of a permeable structure were investigated in the past \cite{Li1994}. In this appendix, we reformulate the existing tube-wave generation model \cite{Li1994} using the expressions that are consistent with the recent literature on open-fracture models \cite<e.g.,>[]{Ionov2007,Bakku2013,Minato2017JGR}. This formulation includes, 1. defining the continuity equation for the fluid volume due to the dynamic change of the layer thickness along with the relation between the fluid flow and the pressure gradient (i.e., Darcy's law), 2. deriving the volume of fluid flowing from the permeable structure (open fracture or porous layer) into the borehole by solving the continuity equation with appropriate boundary conditions, and 3. relating the fluid volume with the pressure amplitudes at the borehole. For a comparison, we will also derive here the tube-wave amplitude based on the same formulation using the diffusion equation considered in this paper (equation \ref{eq:PDE_por}).

We start from the continuity equation and Darcy's law that are considered in an earlier model \cite<see equations 7 and 11 in>[]{Li1994}:
\begin{linenomath*}
 \begin{align}
  -\frac{\partial q(r,\omega)}{\partial r}-\frac{q(r,\omega)}{r}&=-i\omega \Delta L(\omega)-i\omega\frac{L_0}{K_d}\bar{p}(r,\omega), \label{eq:Li_PDE} \\
  q(r,\omega)&=-L_0\frac{\kappa_0}{\eta}\frac{\partial \bar{p}}{\partial r}, \label{eq:Li_Darcy}
 \end{align}
\end{linenomath*}
where $\bar{p}$ indicates the pressure within a porous layer as considered in \citeA{Li1994}, $K_d$ is the drained bulk modulus of the layer, $L_0$ is the static layer thickness, and $\Delta L$ is the dynamic change of the thickness from $L_0$ due to elastic wave propagation. The dynamic thickness change $\Delta L$ can be defined as the difference of the vertical displacement at the upper and lower boundaries of the layer followed by the small-$L_0$ approximation \cite{Li1994}:
 \begin{linenomath*}
 \begin{equation}
\Delta L(\omega)=-k_p^2L_0D_{\rm E},
 \end{equation}
\end{linenomath*}
 where we use our notation of the potential amplitude ($D_{\rm E}$). Note that $K_d$ in equation \ref{eq:Li_PDE} is a free parameter, and one may assign any value which is a function of $\phi$ \cite{Li1994}. In this Appendix, we assume Reuss average, i.e., $K_d^{-1} \approx \phi K_f^{-1}+(1-\phi)K_S^{-1}$, as also suggested in \citeA{Li1994}. Equation \ref{eq:Li_PDE} is solved for $\bar{p}$ with the following boundary condition at the borehole intersection:
\begin{linenomath*}
 \begin{equation}
  \bar{p}(r,\omega)|_{r=R}=\bar{p_t}(\omega), \label{eq:BC_Ionov}
 \end{equation}
\end{linenomath*}
where $\bar{p_t}$ is the generated tube-wave amplitude that we are interested in. This boundary condition (equation \ref{eq:BC_Ionov}) is suggested by \citeA{Ionov2007} and \citeA{Bakku2013}. On the other hand, the original formulation in \citeA{Li1994} considers a constant boundary-value independent of time \cite<see equation 15 in>[]{Li1994}. This condition is the same as that in \citeA{Beydoun1985}, where the tube-wave amplitude is assumed to be small \cite<see the assumption 4 in>[]{Beydoun1985}. Such assumption is not necessary for the boundary condition of equation \ref{eq:BC_Ionov}. Note that the original boundary condition in \citeA{Li1994} can be obtained if we assume $\bar{p}(R,\omega)=0$ ($\omega \ne 0$). Next, using equations \ref{eq:Li_PDE}, \ref{eq:Li_Darcy}, and \ref{eq:BC_Ionov}, the pressure $\bar{p}$ can be solved as
 \begin{linenomath*}
 \begin{equation}
  \bar{p}(r,\omega)=\left(\bar{p_t}-\frac{\Delta L}{L_0}K_d\right)\frac{K_0\left(\sqrt{-i\omega}r/\bar{a}\right)}{K_0\left(\sqrt{-i\omega}R/\bar{a}\right)}+\frac{\Delta L}{L_0}K_d, \label{eq:Li_p}
 \end{equation}
\end{linenomath*}
where
 \begin{linenomath*}
 \begin{equation}
  \bar{a}^2=\frac{\kappa_0 K_d}{\eta}.
 \end{equation}
\end{linenomath*}
The rate of the fluid volume ($\Delta \bar{V}$ $\rm m^3/s$) flowing from the porous layer to the borehole is defined as
 \begin{linenomath*}
 \begin{equation}
  \Delta \bar{V}=-2\pi R q|_{r=R}, \label{eq:Li_dV}
 \end{equation}
\end{linenomath*}
and the fluid volume is related to the generated tube-wave amplitude \cite<e.g.,>[]{Ionov2007,Bakku2013} as
 \begin{linenomath*}
 \begin{equation}
  \bar{p_t}=\frac{\rho_f C_T}{2\pi R^2}\Delta \bar{V}. \label{eq:Li_dV2pt}
 \end{equation}
\end{linenomath*}
Finally, using equations \ref{eq:Li_Darcy}, \ref{eq:Li_p}, \ref{eq:Li_dV}, and \ref{eq:Li_dV2pt}, the tube-wave amplitude can be obtained as
 \begin{linenomath*}
 \begin{equation}
  \bar{p_t}=\frac{\rho_f C_T\frac{L_0}{R} \frac{\kappa_0}{\eta} \frac{\sqrt{-i\omega}}{\bar{a}}K_1\left(\sqrt{-i\omega}r/\bar{a}\right)/K_0\left(\sqrt{-i\omega}R/\bar{a}\right)}{1+\rho_f C_T\frac{L_0}{R} \frac{\kappa_0}{\eta} \frac{\sqrt{-i\omega}}{\bar{a}}K_1\left(\sqrt{-i\omega}r/\bar{a}\right)/K_0\left(\sqrt{-i\omega}R/\bar{a}\right)}\frac{K_d\Delta L}{L_0}. \label{eq:Li_pt}
 \end{equation}
\end{linenomath*}
Equation \ref{eq:Li_pt} is the effective-source model of the tube-wave amplitude based on \citeA{Li1994}.

Similar to the derivation above, one can derive the effective-source model based on the theory developed in this study. This is achieved by considering $p=\tilde{p_t}$ in equation \ref{eq:p_por} and using equations \ref{eq:Li_Darcy}, \ref{eq:Li_dV}, and \ref{eq:Li_dV2pt}:
 \begin{linenomath*}
 \begin{equation}
  \tilde{p_t}=\frac{\rho_f C_T\frac{L_0}{R} \frac{\kappa_0}{\eta} \frac{\sqrt{-i\omega}}{a}K_1\left(\sqrt{-i\omega}r/a\right)/K_0\left(\sqrt{-i\omega}R/a\right)}{1+\rho_f C_T\frac{L_0}{R} \frac{\kappa_0}{\eta} \frac{\sqrt{-i\omega}}{a}K_1\left(\sqrt{-i\omega}r/a\right)/K_0\left(\sqrt{-i\omega}R/a\right)}p^{\rm ext}_{\rm por}. \label{eq:Ionov_pt}
 \end{equation}
\end{linenomath*}
The definition of $p^{\rm ext}_{\rm por}$ can be found in equation \ref{eq:pext}.

\section{The field source wavelet and Q analyses of synthetic data}
\label{Append:Source_Q}

\begin{figure}
\centering
 \noindent\includegraphics{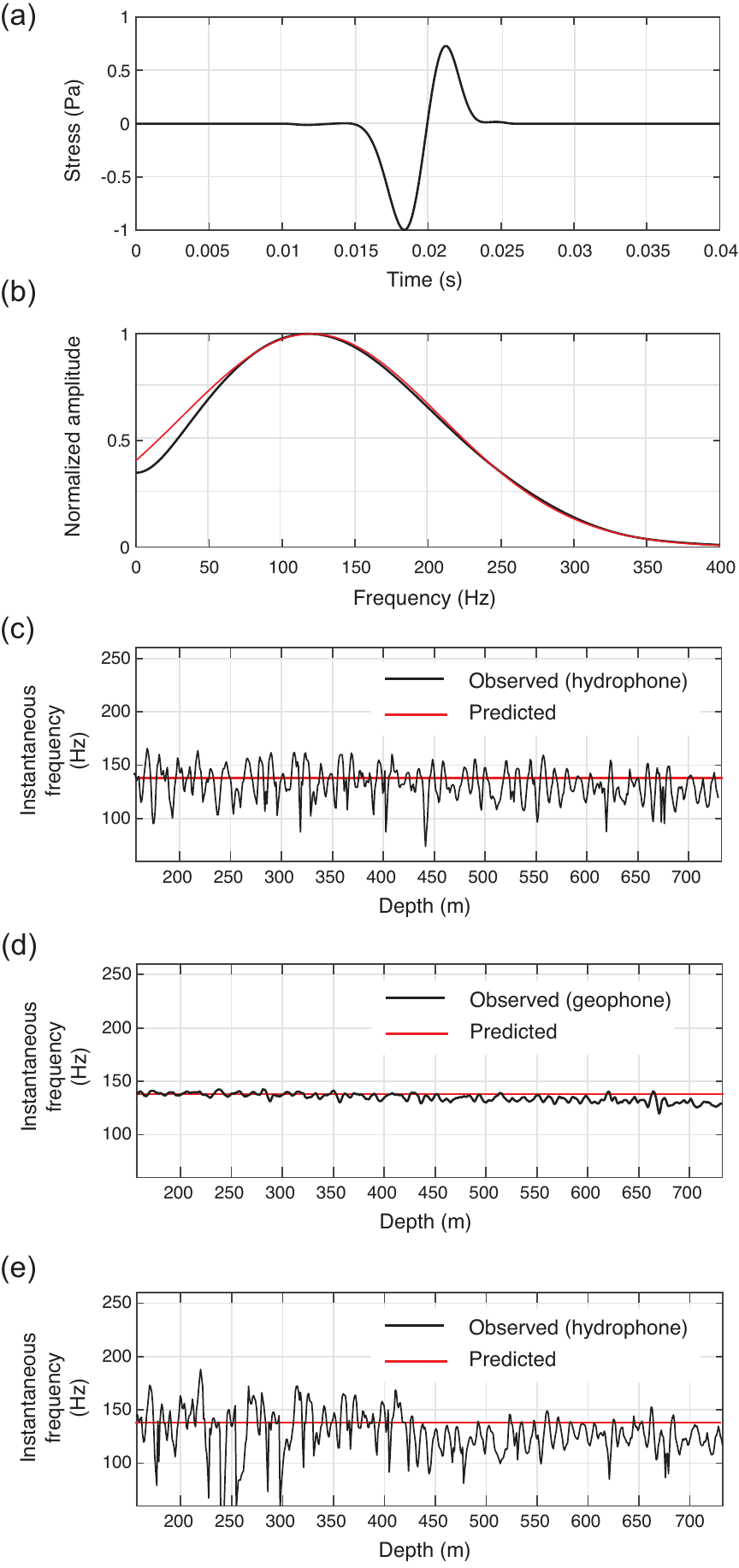}
\caption{(a) Source signature estimated from field data, which is also assumed in calculating the pressure response (hydrophone data). (b) The frequency spectrum of the source signature (black line), and that assumed in calculating the centroid-frequency shift in the attenuation analyses (red line), see Section \ref{Sec:Field_Preproc}. (c) Instantaneous frequency at the envelope maxima of the direct waves in the calculated waveforms shown in Figure \ref{fig:image_elastic}(c). The red line indicates the expected frequency shift (no shift due to $Q$=$\infty$). (d) Same as (c) but using the geophone waveforms (vertical particle velocity at the elastic formation) calculated using the propagator matrix method (\ref{Append:phiE}). (e) Same as (c) but using the calculated waveforms shown in Figure \ref{fig:image_perm}(b).}
\label{fig:Source_Q}
\end{figure}

The wavelet of the seismic source in the field hydrophone data is estimated from data recorded between 589 m and 606 m depth (18 traces) by aligning the direct waves using the picked travel times, then calculating the averaged waveform to mitigate the interferences of tube waves, and finally time-windowing the waveform. Figure \ref{fig:Source_Q}(a) shows the estimated source wavelet. The same wavelet is also utilized as an input stress component of the incident P wave ($\sigma_{zz}$) in modeling the synthetic data (Section \ref{Sec:Field_Synth}). 

Figure \ref{fig:Source_Q}(b) shows the amplitude spectrum of the estimated source wavelet (black line). The red line in Figure \ref{fig:Source_Q}(c) shows the assumed Gaussian spectrum (centroid frequency of 120 Hz and the variance of 8$\times 10^{3}$ $\rm Hz^2$) in calculating the frequency shift in Figure \ref{fig:Q}(b). The frequency shift is calculated using equations (3) and (11b) of \citeA{Quan1997} for a two-layer velocity model for which the travel times are marked by the yellow dashed line in Figure \ref{fig:Q}(a).

Next, in order to check a validity of the attenuation analyses of \citeA{Vesnaver2020} for borehole hydrophone data, we apply the same procedure as in Section \ref{Sec:Field_Preproc} to synthetic data. The black line in Figure \ref{fig:Source_Q}(c) shows the picked instantaneous frequency at the envelope maxima of the direct wave of the modeled pressure data shown in Figure \ref{fig:image_elastic}(c). The theoretical prediction of the centroid frequency (red line in \ref{fig:Source_Q}c) is calculated from the black line of Figure \ref{fig:Source_Q}(b) and with $Q^{-1}=0$ in the forward modeling (i.e., no frequency shifting). The estimated instantaneous frequency shows a large fluctuation due to the interferences of tube waves with the direct P wave, as it is clear from the tiny fluctuations that are observed in synthetic geophone data, i.e., vertical motion at the borehole wall (Figure \ref{fig:Source_Q}d). Furthermore, the inclusion of porous layers in the pressure response increases the noise in the picked frequency (Figure \ref{fig:Source_Q}e). Nevertheless, the observed fluctuations in the instantaneous frequencies are quasi-random around the true values (Figure \ref{fig:Source_Q}d and \ref{fig:Source_Q}e). They do not explain the long-wavelength trend in the field data (Figure \ref{fig:Q}b). These results, therefore, suggest that the frequency shift observed in the field data (Figure \ref{fig:Q}b) is more likely due to attenuation of P waves than artificial effects due to the interferences of tube waves. 

\newpage

\acknowledgments

\textbf{Author Contributions}. S.M.: Conceptualization, Methodology, Formal Analysis, Visualization, Software, Data Curation, Writing -- original draft; T.K.: Investigation, Data Curation; R.G.: Writing -- review and editing.

\textbf{Acknowledgements}. The work of S.M. has received financial support from OYO corporation, Japan.


\begingroup
\raggedright
\bibliography{mybib02}
\endgroup

\end{document}